\documentclass[a4paper,fleqn]{cas-sc}

\usepackage[authoryear]{natbib}
\usepackage{amsmath}
\usepackage{amssymb}
\usepackage{array}
\usepackage{booktabs}
\usepackage{graphicx}
\usepackage{float}
\usepackage[labelfont=bf,singlelinecheck=false,justification=raggedright]{caption}
\usepackage[T1]{fontenc}
\makeatletter
\newenvironment{plainfigurefloat}[1][!htbp]{\@float{figure}[#1]}{\end@float}
\makeatother
\usepackage{tabularx}
\usepackage{multirow}
\newcolumntype{Y}{>{\raggedright\arraybackslash}X}
\usepackage{enumitem}
\ExplSyntaxOn
\RenewDocumentCommand \printemails { }
  {
    \group_begin:
    \int_compare:nNnTF { \int_use:N \g_ead_int } > { 0 }
      {
        \par \vskip6pt
        {\raggedright
        \int_compare:nTF { \g_ead_int = 1 }
          { \textit{Email~address:\c_space_token} }
          { \textit{Email~addresses:\c_space_token} }
        \seq_use:Nn \g_stm_ead_seq { ;~ }
        \par}
      }
      { }
    \group_end:
  }
\ExplSyntaxOff

\def\tsc#1{\csdef{#1}{\textsc{\lowercase{#1}}\xspace}}
\tsc{WGM}
\tsc{QE}

\ExplSyntaxOn
\cs_gset:Npn \__first_footerline:
{
	\group_begin:
	\small
	\sffamily
	\ifnum\theblind>0\relax
	\else
	\__short_authors: :~
	\fi
	{ \rmfamily \itshape Preprint~submitted~to~Knowledge-Based~Systems }
	\group_end:
}
\ExplSyntaxOff

\begin{document}
\let\WriteBookmarks\relax
\def\floatpagepagefraction{1}
\def\textpagefraction{.001}

\shorttitle{CF-RL-TOPSIS Fusion for Talent Recommendation}
\shortauthors{Canay}
\title[mode=title]{An Interpretable CF-RL-TOPSIS Fusion Model for Skills-Aware Talent Recommendation}

\author[1]{{\"O}zkan Canay}[orcid=0000-0001-7539-6001]
\ead{canay@sakarya.edu.tr}
\corref{cor1}
\affiliation[1]{organization={Sakarya University},
               addressline={Department of Information Systems and Technologies},
               addressline={Faculty of Computer and Information Sciences},
               city={Sakarya},
               country={T\"urkiye}}
\cortext[cor1]{Corresponding author}

\begin{abstract}
Effective skills-aware talent recommendation must balance behavioral transition patterns, trajectory-sensitive adaptation, and occupation-level criteria that can be inspected by human decision makers. Evidence from public benchmarks on how these signals interact, however, remains limited. This study proposes CF-RL-TOPSIS, an interpretable late-fusion model that integrates a transition-aware collaborative branch, a compact reinforcement-style occupation-family bandit, and an entropy-weighted Technique for Order Preference by Similarity to Ideal Solution branch constructed from six semantic proxies; the validation-selected fusion coefficients remain directly auditable. The model is evaluated on two frozen public information and communication technology talent-history benchmarks, JobHop and Karrierewege, using repeated chronological top-5 ranking and paired Wilcoxon tests. On JobHop the full hybrid attains NDCG@5 = 0.3040 +/- 0.0073 and significantly surpasses repeat-last, item Markov, transition-aware collaborative filtering, the CF+TOPSIS hybrid, GRU4Rec, and SASRec (p <= 0.0039 across planned comparisons). On Karrierewege the hybrid remains competitive but does not significantly exceed the strongest Markov baseline, revealing a persistence-dominated setting in which the bandit branch appropriately shrinks to near-zero weight. Proxy-sensitivity, family-level deep Q-network, and runtime checks support this interpretation, and a worked user-level case shows how branch scores, criterion weights, and rank shifts can be inspected for an individual recommendation. The contribution is therefore not a benchmark-agnostic superiority claim, but a reproducible account of the conditions under which transparent late fusion adds value beyond simple continuation heuristics. In semantically rich, non-saturating talent-history regimes the three branches reinforce one another; in persistence-dominated regimes the same architecture remains competitive through its collaborative backbone, with the adaptive branch correctly inactive.
\end{abstract}

\maketitle

\section{Introduction}\label{sec:introduction}

Technology talent recommendation sits at the intersection of recommender systems, sequential prediction, labor-market intelligence, and decision support. Unlike entertainment recommendation, the output of a talent recommender can influence career exploration, upskilling, hiring, and internal mobility decisions. In realistic settings, a useful recommender is rarely expected to recover only the statistically most likely next occupation. It must also remain aligned with interpretable occupational properties such as skill breadth, digital intensity, mobility, and role level while still adapting to an individual's observed trajectory \citep{li2024recent, zheng2022survey, caromartinez2021explainable}. This combination of requirements makes the problem structurally hybrid from the outset.

This structural hybridity creates a clear methodological tension. Collaborative approaches are effective at exploiting shared regularities in transition histories, but they do not by themselves provide a criterion-level rationale \citep{burke2002hybrid, chen2020attributeaware}. Sequential and reinforcement-style models promise adaptation to evolving user trajectories, yet their contribution can be difficult to interpret when public histories are short, sparse, or strongly dominated by repetition \citep{shani2005mdp, tang2019reinforcement, lin2023survey}. Multi-criteria decision-making methods such as Technique for Order Preference by Similarity to Ideal Solution (TOPSIS) are attractive because they retain an explicit ideal-versus-anti-ideal ranking logic, but they are not designed to replace collaborative evidence where historical behavioral structure remains central \citep{albashiri2018topsiscf, monti2021systematic, mukhametzyanov2021specific}.

The broader recommendation literature has long emphasized that no single model family dominates across all data regimes \citep{burke2002hybrid, roy2022systematic, li2024recent}, and the same lesson appears in job, person--job, and career recommendation research. Prior studies have distinguished explicit and implicit information regimes \citep{reusens2017explicit}, extracted implicit skills from resumes and job descriptions \citep{gugnani2020skills}, and explored content-based and collaborative combinations for job matching \citep{dhameliya2019job, alsaif2022matchedjob}, while contextual and talent-acquisition-oriented matching systems represent another active line \citep{channabasamma2022talent, azri2025contextjob}. More recent work has extended the field toward large language model (LLM)-enhanced resume completion and graph-based recruitment reasoning \citep{du2024llmjob, wu2024llmgraphjob}, while large-scale public-employment and explainable multi-stakeholder recommendation have begun to receive direct attention \citep{bied2023toward, schellingerhout2024explainablejob, tang2025explainablepjr}. Taken together, these studies suggest that talent recommendation should not be treated as a one-signal problem.

The benchmark question is equally important, since public talent-history datasets remain far fewer and structurally more heterogeneous than the benchmark suites available in mainstream recommendation domains. Some public datasets are primarily text-matching corpora, while others are trajectory datasets whose sequence dynamics can make simple persistence baselines unexpectedly strong \citep{poncio2024systematicjob, senger2024karrierewege, senger2025realistic}. Consequently, a model that appears strong on one public talent-history dataset may reflect the benchmark regime itself rather than a generally useful fusion of signals.

Against that background, the paper introduces an interpretable collaborative-reinforcement multi-criteria score-fusion architecture for skills-aware technology talent recommendation. The proposed model combines three branches within a strict late-fusion pipeline: a transition-aware collaborative filtering branch, a lightweight reinforcement-style bandit branch that models occupation-family transitions, and an entropy-weighted TOPSIS branch derived from occupation-side semantic proxies. Interpretability here refers to model transparency at design level: the three branch scores remain separate, the fusion coefficients are made directly inspectable, and the TOPSIS component preserves a criterion-level rationale instead of relying on a post-hoc explanation layer \citep{zhang2020explainable, tang2025explainablepjr}. Rather than presenting reinforcement learning as a standalone solution to talent recommendation, the study tests whether a disciplined fusion of collaborative, adaptive, and multi-criteria signals yields measurable benefit under fixed chronological evaluation while preserving branch-level interpretability.

The empirical design is therefore deliberately benchmark-sensitive. Rather than relying on a single dataset, this study uses two public information and communication technology (ICT)-focused talent-history benchmarks with different dynamics. JobHop is used as the primary benchmark because it yields the clearest evidence for the proposed hybrid. Karrierewege is retained as a secondary robustness benchmark because of its public visibility, European Skills, Competences, Qualifications and Occupations (ESCO) linkage, and importance as a public career-path testbed, even though it exhibits much stronger persistence behavior. This dual-benchmark design allows the study to answer not only whether the hybrid works, but also under what conditions its three branches remain active and when the problem collapses toward simpler sequential structure \citep{bied2023toward, senger2024karrierewege, senger2025realistic}.

Within this dual-benchmark design, the retained occupation universe is also restricted to a documented ICT/software subset, keeping the study centered on software engineering, infrastructure, cybersecurity, data and artificial intelligence (AI), digital-product, and related technology roles. This focus aligns the contribution with skills-aware recommendation, benchmark design, sequential modeling, and interpretable recommendation modeling rather than allowing it to drift into a generic HR analytics narrative \citep{sun2021skillrl, sun2025marketaware, wu2024llmgraphjob}.

Three questions guide the empirical inquiry:
\begin{itemize}[label={}, leftmargin=1.5em]
	\item \textit{RQ1:} Does the proposed CF+RL+TOPSIS score-fusion model outperform strong sequential, collaborative, and pairwise-hybrid baselines on a primary public talent-history benchmark?
	\item \textit{RQ2:} Does the reinforcement-style branch provide additive value beyond CF+TOPSIS, or does its contribution collapse when the benchmark is dominated by role persistence?
	\item \textit{RQ3:} Can a dual-benchmark design provide a more reliable empirical assessment of talent recommendation performance than a single-benchmark evaluation?
\end{itemize}

The study makes three primary contributions:

\begin{itemize}
	\item It introduces a reproducible ICT-focused dual-benchmark setup for public talent-history data, with JobHop as the primary benchmark and Karrierewege as the secondary robustness benchmark.
	\item It proposes an interpretable CF+RL+TOPSIS late-fusion architecture for talent recommendation in which each signal family can be tuned, ablated, and inspected independently.
	\item It provides a benchmark-sensitive empirical analysis showing when reinforcement-style modeling remains meaningfully active and when the recommendation problem collapses toward simpler persistence-dominated sequential structure.
\end{itemize}

The remainder of the paper is organized as follows. Section~\ref{sec:related} reviews the recent literature on job recommendation, career-path prediction, hybrid recommendation, and benchmark validity. Section~\ref{sec:method} presents the end-to-end study workflow, benchmark construction protocol, and the proposed CF-RL-TOPSIS fusion model. Section~\ref{sec:results} reports the primary and secondary benchmark results together with the ablation analysis and cross-benchmark comparison. Section~\ref{sec:discussion} interprets the empirical findings, clarifies the paper's originality claim, and discusses limitations and practical implications. Finally, Section~\ref{sec:conclusion} concludes the paper and outlines the most important directions for future work.

\section{Related Work}\label{sec:related}

Recent talent recommendation research now spans classical matching, career-sequence prediction, explainability, and benchmark design. The literature reviewed below is selected to position the paper against those neighboring lines of work rather than to exhaust the broader recommender-systems field.

\subsection{Job, Person--Job, and Talent Recommendation}

Job and talent recommendation have matured into a distinct branch of recommender systems with their own data, explainability, and evaluation challenges. Early work in the area highlighted the difference between explicit and implicit information regimes \citep{reusens2017explicit}, while later studies emphasized that job recommendation is often inseparable from skill extraction and representation learning over resumes and job descriptions \citep{gugnani2020skills}. Practical systems still rely heavily on content-based, collaborative, and hybrid matching strategies, but their relative strengths depend strongly on data availability, sparsity, and the role of textual side information \citep{dhameliya2019job, alsaif2022matchedjob, channabasamma2022talent}.

Recent work confirms that this area remains methodologically active: \citet{bied2023toward} describe job recommendation for public employment services under extreme interaction sparsity and mandatory scalability constraints. \citet{du2024llmjob} and \citet{wu2024llmgraphjob} show how LLMs can be used to enrich resumes or better understand recruitment-graph structure. \citet{guan2024jobformer} propose a skill-aware Transformer architecture for job-description-to-user matching, which is methodologically related but assumes access to job-description text and user-skill profiles rather than the public sequence-only career histories used here. \citet{hoque2026fuzzytopsis} apply a fuzzy TOPSIS layer over LLM-derived candidate attributes for personnel selection, illustrating that multi-criteria decision rules remain actively combined with modern representation learning in talent-matching tasks. \citet{schellingerhout2024explainablejob} foregrounds explainability and multi-stakeholder requirements, arguing that candidate-facing recommendation needs trust and traceability in addition to ranking quality. These developments collectively show that current job recommendation research is no longer confined to plain text matching or collaborative filtering.

Surveys also suggest that the field remains fragmented: \citet{poncio2024systematicjob} reviews internship and job recommender techniques and finds a continuing dominance of profile matching, collaborative filtering, and hybrid ranking methods with limited interpretability emphasis. \citet{zou2024reviewjob} similarly describe a broad but methodologically uneven design space. This fragmentation matters here because claims about model superiority in talent recommendation remain tightly coupled to data regime, benchmark design, and the role of side information.

\subsection{Career-Path Prediction and Public Benchmarks}

Career-path prediction is closely related to job recommendation, but it foregrounds sequential occupational histories and next-step forecasting rather than one-shot document matching. This subfield has recently gained momentum because public benchmark availability has improved. The Karrierewege dataset line and its follow-up evaluation study are especially important because they elevate public benchmark construction, realistic evaluation, and cross-scenario comparison to first-class research problems \citep{senger2024karrierewege, senger2025realistic}. In particular, \citet{senger2025realistic} show that model conclusions can change materially with evaluation scenario and grounding strategy, which reinforces the view that benchmark selection is itself a substantive methodological variable. More recent work continues to push sequential career-path modeling forward: \citet{cui2026scp} couple a knowledge-graph backbone with a large language model to predict the next career hop, and \citet{feng2026personjob} represent person--job fit through multi-temporal career trajectory modeling. Both confirm that sequential occupational structure remains a productive modeling target, but they rely on richer text or graph side information than the public sequence-only career histories used here.

Public-employment recommendation research reaches a similar conclusion from a different angle. \citet{bied2023toward} demonstrate that large-scale job recommendation can be severely constrained by data sparsity, privacy, and operational scalability. These constraints make the absence of strong public benchmarks more than a convenience problem: without benchmark discipline, it becomes difficult to know whether a measured gain reflects a substantive model contribution or an unusually favorable dataset regime. This point is central to the dual-benchmark design adopted here.

Closer to the occupation-transition framing of the present paper, \citet{zha2024careermobility} model job-title transitions through uncertainty-aware graph autoencoders for career-mobility analysis. That line of work underscores the value of explicitly modeling transition structure, but it does not address the complementarity question between collaborative, adaptive, and multi-criteria signals that motivates the architecture studied here.

\begin{table*}[width=\FullWidth,pos=!htbp]
	\centering
	\caption{Selected recent studies in job, talent, and career-path recommendation.}
	\label{tab:recent_job_talent_work}
	\footnotesize
	\renewcommand{\arraystretch}{1.3}
	\begin{tabularx}{\textwidth}{@{} 
			>{\raggedright\arraybackslash\hsize=0.6\hsize}X 
			>{\raggedright\arraybackslash\hsize=0.8\hsize}X 
			>{\raggedright\arraybackslash\hsize=1.2\hsize}X 
			>{\raggedright\arraybackslash\hsize=1.4\hsize}X 
			@{}}
		\toprule
		\textbf{Study} & \textbf{Method} & \textbf{Technique(s)} & \textbf{Findings} \\
		\midrule
		\cite{alsaif2022matchedjob} & Job-profile representation matching & Matched representation learning; ontology-aware profiles; neural user--job embeddings & Evaluated on job-posting data from 3 Saudi cities and reported accuracy, precision, recall, and F1. \\
		
		\cite{bied2023toward} & Public-employment recommendation at scale & Faceted ranking; sparsity-aware retrieval; scalable serving & Targets millions of job seekers and hundreds of thousands of ads while reporting an approximately 2-orders-of-magnitude inference-time gain at comparable recall. \\
		
		\cite{channabasamma2022talent} & Contextual talent acquisition & Context-aware recommendation; NLP resume parsing; job--profile matching & Manual verification in the paper confirms that 67 of 100 candidates in the inspection set meet the minimum skill threshold for shortlisting. \\
		
		\cite{poncio2024systematicjob} & Systematic review of internship/job recommenders & Structured literature review; narrative synthesis; profile-matching taxonomy & Organizes the review around 3 research questions and 4 systematic-review phases. \\
		
		\cite{zou2024reviewjob} & Survey of job recommender-system designs & Taxonomy of content-based, collaborative, knowledge-based, and hybrid/deep variants & Synthesizes the field around 4 recurrent technique families used in current job recommenders. \\
		
		\cite{du2024llmjob} & LLM-enhanced job recommendation & LLM resume completion; GAN refinement; hybrid matching & Evaluated on 3 real recruitment datasets and reports relative gains of 6.65\%, 7.40\%, and 6.42\% over the strongest baselines. \\
		
		\cite{wu2024llmgraphjob} & Graph-based online job recommendation & LLM graph understanding; LoRA fine-tuning; graph-structured prompting & Evaluated on 2 real-world recruitment datasets and reports 2 main findings about path weighting and prompt-position bias. \\
		
		\cite{sun2021skillrl} & Long-term job-skill recommendation & Deep reinforcement learning; interpretable policy learning; cost-aware optimization & Optimizes 20-step job-skill recommendation paths on a real IT-oriented dataset while balancing utility, interpretability, and learning cost. \\
		
		\cite{senger2024karrierewege} & Public career-path benchmark construction & Benchmark curation; ESCO-linked occupation graphing; career-history sequencing & Introduces more than 500,000 career paths linked to over 3,000 occupations, nearly 14,000 skills, and 28 languages. \\
		
		\cite{senger2025realistic} & Realistic career-path prediction (CPP) evaluation study & Multi-scenario evaluation; grounding analysis; model-family comparison & Compares 4 model families on 2 public datasets and adds a Karrierewege+ scenario derived from 100,000 resumes. \\
		
		\cite{schellingerhout2024explainablejob} & Explainable multi-stakeholder job recommendation & Stakeholder-specific explanations; knowledge-graph reasoning; qualitative elicitation & Frames the problem around 3 stakeholder groups and reports an initial user study with 6 participants. \\
		
		\cite{tang2025explainablepjr} & Systematic review of explainable person--job recommendation & Explainable artificial intelligence (XAI) taxonomy; comparative review; person--job recommendation (PJR)-focused synthesis & Reviews 85 studies selected from 150 screened articles published between 2019 and August 2025. \\

		\cite{cui2026scp} & LLM + knowledge-graph sequential job-mobility prediction & Career knowledge graph; sub-graph retrieval; multi-modal text--graph adapter; LLM next-hop reasoning & Evaluated on two real-world career datasets and reports consistent gains over neural and embedding-based job-mobility baselines on next-hop prediction. \\

		\cite{hoque2026fuzzytopsis} & LLM-driven personnel selection with multi-criteria ranking & Fine-tuned DistilRoBERTa attribute scoring; Fuzzy TOPSIS aggregation; ranking with attribute confidences & Reaches up to 91\% accuracy on personnel attributes (Experience, Overall) using a custom LinkedIn-derived ranking benchmark. \\
		\bottomrule
	\end{tabularx}
\end{table*}

Table~\ref{tab:recent_job_talent_work} summarizes 14 representative recent studies across public job recommendation, talent acquisition, explainable person--job recommendation, skill recommendation, and career-path prediction. To make the comparison more informative for model design, the table separates each study's high-level methodological objective from its concrete technique stack and reports one transferable headline finding drawn from the abstract, benchmark description, or explicitly stated study setup. Three patterns are especially visible: recent work increasingly combines representation learning with behavioral or graph signals, explainability is becoming a visible design target, and benchmark realism remains an unresolved issue despite rapid model innovation. What remains comparatively thin is evidence on interpretable late-fusion designs evaluated under public, benchmark-aware, sequential protocols, which is the gap targeted here.

\subsection{Hybrid Recommendation, Explainability, and Multi-Criteria Ranking}

Hybrid recommendation remains the most natural umbrella for this problem. The classical hybrid-recommender rationale is that collaborative, content-based, and rule-driven signals fail in different ways, so combining them can be more robust than relying on a single model family \citep{burke2002hybrid}. Subsequent surveys reinforce this point, extending it to explainable and knowledge-enhanced settings \citep{roy2022systematic, li2024recent, alhasan2024survey, tiwary2024review, abdulhussien2021behavior}. In job recommendation, the case for hybridization is arguably stronger than in many consumer domains because the system must simultaneously model textual semantics, behavior histories, occupational structure, and stakeholder-facing transparency.

Within this broader hybrid space, pairwise neighbors to the proposed design already exist. Collaborative filtering has been combined with TOPSIS and other multi-criteria decision tools in recommendation more generally \citep{albashiri2018topsiscf, monti2021systematic, baczkiewicz2021mcdm}. Reinforcement-enhanced recommenders have also been combined with collaborative or graph-based components in several application domains \citep{peng2024drlcf, zhou2020interactive, xin2020selfrl, zhao2021dear}, and more recent work such as \citet{liu2026cfrlhybrid} fuses collaborative filtering, matrix factorization, and reinforcement-learning components within a single hybrid framework, although on a domain (e-commerce) distinct from career-history data. Explainable-recommendation research, meanwhile, increasingly stresses that transparent rationale is not optional when recommendations affect consequential decisions \citep{caromartinez2021explainable, schellingerhout2024explainablejob, tang2025explainablepjr}.

This work differs from these lines in both domain and evidence structure: it studies skills-aware technology talent recommendation and evaluates a three-branch late-fusion model under a dual-benchmark protocol rather than presenting a single-dataset hybrid.

The multi-criteria branch deserves separate emphasis. TOPSIS remains attractive because it preserves an interpretable ideal-versus-anti-ideal ranking principle while allowing flexible weighting and explicit criterion inspection \citep{zavadskas2016development, panda2018topsis, mukhametzyanov2021specific}. In talent recommendation, this matters because a recommended occupation is not merely a latent label. It is also a role with inspectable skill, level, digital-intensity, and mobility characteristics that users and institutions may reasonably want to understand.

\subsection{Reinforcement Learning and Skills-Aware Talent Modeling}

The RL literature is relevant here, but it needs careful interpretation. Reinforcement learning has been widely studied for recommendation as sequential decision making \citep{shani2005mdp, intayoad2018reinforcement, tang2019reinforcement, lin2023survey, bangari2021review}. Deep RL variants have shown value in domains such as news, advertising, service recommendation, and interactive recommendation, especially when long-term feedback or exploration--exploitation trade-offs matter \citep{munemasa2018deep, xin2020selfrl, zhao2021dear, zhang2023service, fu2022deep}. However, many of these systems assume richer online logs, denser exposure records, or policy-learning conditions than public talent-history benchmarks can naturally provide.

In the talent domain specifically, the most relevant RL-adjacent studies have concentrated on job-skill recommendation rather than next-occupation ranking. \citet{sun2021skillrl} and \citet{sun2025marketaware} show that explainable deep RL can be useful when the task is to recommend long-term skill acquisition under market and utility constraints. These studies are highly relevant to the paper's motivation because they validate labor-market adaptation as an RL problem. At the same time, their task differs from the task studied here: the present work ranks next occupations from public historical career traces rather than recommending a sequence of skills under a richer long-horizon simulator. This difference explains why the RL branch used here is intentionally lightweight and benchmark-sensitive. Methodologically closer to the explicit reinforcement-learning and multi-criteria combination explored in this paper, \citet{nematollahi2026maudrl} introduce a multi-attribute utility deep reinforcement learning method for sequential multi-criteria decision problems applied to human-resource planning. That work targets a resource-allocation policy rather than top-K occupation ranking, but it confirms that the joint use of reinforcement learning and multi-criteria utility remains methodologically active in 2026 and reinforces the value of an explicit, interpretable late-fusion design in the present setting.

\subsection{Offline Evaluation and Benchmark Validity}

The offline-evaluation literature cautions against treating metric values as benchmark-independent facts. \citet{zhao2022offlineeval} show that offline top-$N$ results depend strongly on evaluation protocol. \citet{jadidinejad2021simpson} demonstrate how offline aggregation can mislead under exposure bias, and \citet{carraro2021debiasing} discuss missing-not-at-random effects and debiasing approaches. These concerns are especially relevant in talent recommendation, where public benchmarks are derived from historical traces rather than controlled exposure logs and where sequence persistence can make simple baselines unexpectedly strong \citep{senger2025realistic}.

The study responds to that literature along three coordinated lines. Both benchmarks are frozen after preparation so that later model iterations cannot silently alter the data regime, while the evaluation itself is chronological and repeated, which separates development-time tuning from robustness reporting. Beyond these protocol choices, the empirical interpretation explicitly centers cross-benchmark differences rather than assuming that a gain observed on one public talent-history dataset should transfer unchanged to all others.

\section{Methodology}\label{sec:method}

The empirical pipeline was designed to keep benchmark construction, model comparison, and interpretation clearly separable. The next subsections follow that same logic, moving from frozen benchmark preparation to model design and then to the evaluation protocol.

\subsection{Overall Workflow}\label{sec:workflow}

Figure~\ref{fig:talent_benchmark_pipeline} summarizes the overall study design from public talent-history sources to cross-benchmark interpretation. The workflow begins with the two public career-history sources, applies an explicit ICT/software occupation audit, constructs chronological user--occupation sequences, prepares occupation-side item tables, and freezes every benchmark artifact needed for later experiments. Both the baseline suite and the proposed CF+RL+TOPSIS model are then evaluated only on those frozen packages, followed by repeated chronological evaluation and cross-benchmark interpretation. This separation is intentional: benchmark preparation is completed once, while model iteration happens only on the frozen artifacts. As a result, later tuning steps cannot silently change the data regime.

\begin{plainfigurefloat}[!htbp]
\begin{minipage}{\textwidth}
	\centering
	\includegraphics[width=\textwidth]{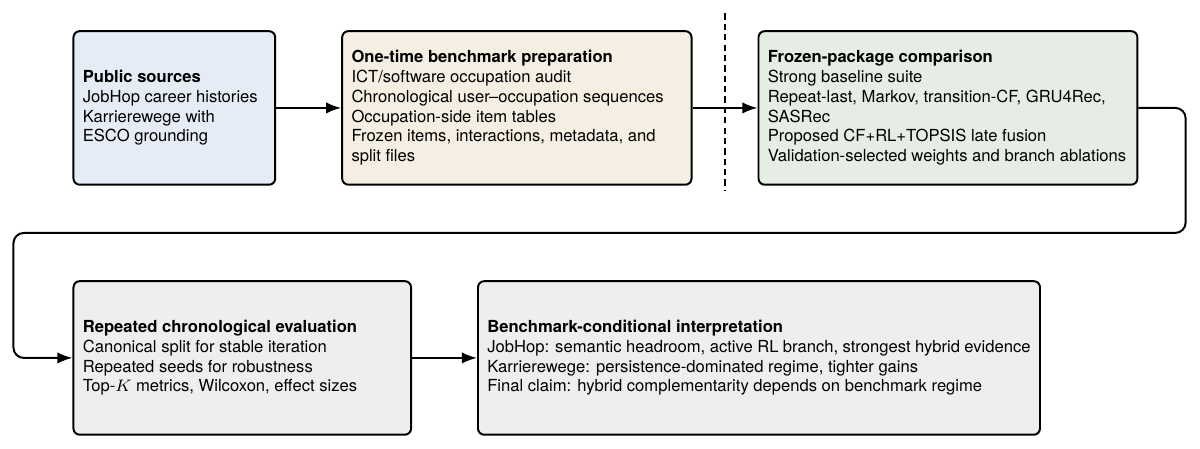}
	\captionof{figure}{End-to-end study design from public talent-history sources to repeated chronological evaluation.}
	\label{fig:talent_benchmark_pipeline}
\end{minipage}
\end{plainfigurefloat}

The same workflow also clarifies the role of the two benchmarks. JobHop is positioned as the primary benchmark because it offers the clearest setting in which collaborative, adaptive, and criteria-based signals can all remain active. Karrierewege is retained as the secondary benchmark because it is a credible public testbed with ESCO grounding and stronger persistence dynamics. Using the same preparation logic for both datasets makes cross-benchmark differences interpretable as substantive benchmark effects rather than preprocessing drift. Figure~\ref{fig:talent_method_pipeline} then details the proposed CF-RL-TOPSIS architecture within that broader study design.

\subsection{Benchmark Construction and Dataset Selection}\label{sec:dataset}

The empirical study is organized around two public talent-history benchmarks rather than a single dataset. This is a deliberate design choice. A single-dataset study would leave the contribution vulnerable to the criticism that the reported gains may be highly benchmark-specific, especially in a domain where public job and career recommendation datasets remain limited and structurally heterogeneous \citep{bied2023toward, dhameliya2019job, senger2025realistic}. The study therefore adopts a primary-secondary benchmark design: JobHop is used as the main benchmark for model comparison, while Karrierewege is retained as a secondary robustness benchmark.

The two datasets belong to the same broad problem family but expose different empirical regimes. JobHop provides semantically rich occupation labels and descriptions derived from anonymized employment histories \citep{johary2025jobhop}, while Karrierewege provides a public ESCO-linked career-path benchmark with strong benchmark legitimacy through its recent public release and follow-up evaluation work \citep{senger2024karrierewege, senger2025realistic}. Both datasets can be recast as next-occupation recommendation tasks, but they differ materially in how strongly simple role persistence dominates the target. JobHop is accessed from its public Hugging Face release under a CC BY 4.0 license, whereas Karrierewege is accessed from the authors' public release under the providers' published terms; the exact source URLs, snapshot references, and retrieval steps are preserved in the replication package.

For both sources, an ICT/software occupation subset is first defined through a documented allow-list and a reproducible audit pass. The retained occupation universe covers software engineering, systems and infrastructure, data and AI, security, hardware-oriented digital roles, digital experience roles, and technology management occupations that remain recognizably inside the computer and information engineering domain. The allow-list is encoded in the benchmark-preparation scripts and preserved in the frozen benchmark metadata so that the filtering decision remains inspectable and repeatable. Each dataset is then transformed into a frozen benchmark with four artifact groups: ordered user histories, occupation-level item metadata, split definitions, and benchmark metadata.

The resulting benchmark statistics are reported in Table~\ref{tab:talent_benchmark_summary}. After filtering, the primary JobHop benchmark contains 2778 users, 47 occupations, and 10826 interactions with an average sequence length of 3.897. The secondary Karrierewege benchmark contains 2617 users, 35 occupations, and 10492 interactions with an average sequence length of 4.009. In both cases, only users with at least three retained ICT occupations are kept, and only occupations observed for at least 25 distinct users are retained. This choice keeps the task warm-started while avoiding degenerate singleton occupations.

\begin{table}[pos=!htbp]
	\centering
	\caption{ICT benchmark summary.}
	\label{tab:talent_benchmark_summary}
	\small
	\renewcommand{\arraystretch}{1.2}
	\begin{tabular}{lcc}
		\toprule
		\textbf{Statistic} & \textbf{JobHop} & \textbf{Karrierewege} \\
		\midrule
		Source type & CSV splits & Public parquet \\
		Users & 2,778 & 2,617 \\
		Occupations & 47 & 35 \\
		Interactions & 10,826 & 10,492 \\
		Average sequence length & 3.897 & 4.009 \\
		Minimum sequence length & 3 & 3 \\
		Maximum sequence length & 11 & 23 \\
		ICT audit scan size & 76 & 58 \\
		Minimum item user support & 25 & 25 \\
		Main role in paper & Primary benchmark & Robustness benchmark \\
		\bottomrule
	\end{tabular}
\end{table}

Two contrasts in Table~\ref{tab:talent_benchmark_summary} are worth highlighting before the next step: Karrierewege is the smaller and more concentrated benchmark in terms of retained occupations (35 vs.\ 47) yet carries longer maximum trajectories (23 vs.\ 11), which foreshadows its stronger persistence regime in the empirical results, while JobHop is the larger and slightly shorter-sequence benchmark, leaving more room for non-trivial transition modeling.

The split protocol is chronological. For each user, a canonical development split is retained for stable iteration using a fixed seed of 20260331. In addition, repeated chronological splits are generated for robustness reporting using seeds 100--109. Under the repeated protocol, the test item is sampled from the valid chronological tail of the sequence, the immediately preceding occupation is used for validation, and the remaining prefix is treated as training history. This preserves temporal order and avoids leaking future occupations into the training prefix. All preprocessing artifacts that learn from data, including the popularity prior, the occupation-to-occupation transition matrix, and the criterion matrix used by the TOPSIS branch, are fit only on the training prefix of each split rather than on the full sequence, which preserves test-set isolation.

The two-benchmark design is not intended to average away benchmark differences; it is intended to expose them. JobHop functions as the paper's main empirical testbed because the proposed hybrid demonstrates its clearest gains there. Karrierewege functions as a robustness dataset because it is public, credible, and domain-relevant, yet empirically much more persistence-dominated. That distinction becomes central in the interpretation of the results.

\subsection{Proposed CF-RL-TOPSIS Fusion Model}\label{sec:proposed_model}

Let $\mathcal{U}$ denote the user set and $\mathcal{I}$ the occupation set. For each user $u \in \mathcal{U}$, the observed career history is represented as an ordered sequence of retained ICT occupations. Recommendation is framed as a next-occupation ranking task: given the training prefix of a user's sequence, the model ranks candidate occupations and is evaluated on the held-out next occupation. Throughout the paper, the ``RL'' label refers to this compact reinforcement-style bandit branch rather than to a full online or deep-RL recommender environment.

For a user $u$ and candidate occupation $i$, the proposed model computes three branch scores in a late-fusion setting that keeps the component recommenders separately tunable and interpretable \citep{burke2002hybrid, roy2022systematic, caromartinez2021explainable}. The resulting fusion score is given in Eq.~\eqref{eq:fusion}:
\begin{equation}
S_{u,i} = \lambda_{\mathrm{CF}} s^{\mathrm{CF}}_{u,i}
        + \lambda_{\mathrm{RL}} s^{\mathrm{RL}}_{u,i}
        + \lambda_{\mathrm{T}} s^{\mathrm{T}}_{u,i},
\label{eq:fusion}
\end{equation}
where $\lambda_{\mathrm{CF}}$, $\lambda_{\mathrm{RL}}$, and $\lambda_{\mathrm{T}}$ are non-negative and sum to one. The coefficients are selected on the validation split and then evaluated unchanged on the test split. The model is therefore a strict late-fusion architecture: the three branches are built independently from the same training data, and only the resulting score vectors are fused.

\medskip
\noindent\textit{Transition-aware collaborative branch.}
The collaborative branch models collective occupation-movement regularities. From the training histories, an occupation-to-occupation transition matrix is first built. A simple item Markov baseline uses only the normalized outgoing probabilities of the user's last observed occupation. The collaborative branch extends this by computing cosine similarities over transition rows and aggregating both a last-state Markov component and a recency-weighted history component. In the implementation, the final CF branch score is a normalized mixture of a 0.85-weighted last-item transition vector and a 0.15-weighted history-aware component that itself blends transition and similarity evidence. This branch serves as the structural backbone of the recommender and is consistent with the broader role of collaborative and transition-aware evidence in recommenders built from historical interaction structure \citep{chen2020attributeaware, alsaif2022matchedjob, bied2023toward}.

More explicitly, if $p_{j \rightarrow i}$ denotes the normalized transition score from occupation $j$ to occupation $i$, $\sigma(j,i)$ denotes transition-row cosine similarity, $\rho_t$ denotes a recency weight over the user's history, $h_t$ denotes the $t$-th occupation in the user's training prefix, and $T_u$ denotes its length, then the branch follows the form in Eq.~\eqref{eq:cfbranch}:
\begin{equation}
s^{\mathrm{CF}}_{u,i} = \beta\, \tilde{p}_{\ell_u \rightarrow i}
 + (1-\beta)\left[\gamma \sum_{t=1}^{T_u} \rho_t\, \tilde{p}_{h_t \rightarrow i}
 + (1-\gamma)\sum_{t=1}^{T_u} \rho_t\, \tilde{\sigma}(h_t,i)\right],
\label{eq:cfbranch}
\end{equation}
where $\ell_u$ is the last occupation in the training prefix, tildes denote min-max normalization, and the implementation uses $\beta=0.85$ and $\gamma=0.7$.

\medskip
\noindent\textit{Reinforcement-style branch.}
The RL branch is a compact bandit-style adaptive component rather than a full policy-learning environment. Each occupation is mapped to an occupation family such as software engineering, infrastructure and security, data and AI, digital experience, hardware and automation, or technology management. Training transitions are used to update a family-to-family value table $Q$ using positive reinforcement for observed transitions and weak negative reinforcement for sampled alternative actions. The choice of a compact bandit-style state abstraction follows the general sequential-decision perspective of RL recommenders while remaining compatible with the limited density of public talent-history benchmarks \citep{shani2005mdp, intayoad2018reinforcement, lin2023survey, sun2021skillrl}. The one-step update is summarized in Eq.~\eqref{eq:bandit}:
\begin{equation}
Q(s,a) \leftarrow Q(s,a) + \eta \left(r - Q(s,a)\right),
\label{eq:bandit}
\end{equation}
with $\eta$ denoting the update step size and $r$ denoting either a positive reward for an observed transition or a weak negative reward for a sampled alternative. Because this branch is bandit-style, no bootstrapped future-value term is included. After repeated passes through the observed histories, the resulting family-level values are normalized and converted into occupation-level scores through the family assignment of each candidate occupation. A recency-weighted family-bias term is added so that the branch captures not only the last observed family but also the broader family composition of the user's history. This compact design keeps the method clearly distinct from a full deep-RL environment while still permitting a test of whether a lightweight adaptive signal adds value beyond collaborative and criteria-based evidence.

If $f(i)$ denotes the family of occupation $i$, then the family-level preference vector for user $u$ can be written as
\begin{equation}
g_u = \omega\, Q_{f(\ell_u),\,:} + (1-\omega)\, b_u,
\label{eq:familyvector}
\end{equation}
where $b_u$ is the recency-weighted family-bias vector and $\omega$ is the transition weight. The occupation-level RL score is then given by Eq.~\eqref{eq:rlscore}:
\begin{equation}
s^{\mathrm{RL}}_{u,i} = 0.75\, g_{u,f(i)} + 0.25\, \pi_i,
\label{eq:rlscore}
\end{equation}
where $g_{u,f(i)}$ is the $f(i)$-th component of $g_u$ and $\pi_i$ is the normalized popularity prior of occupation $i$.

\medskip
\noindent\textit{Entropy-weighted TOPSIS branch.}
The multi-criteria branch uses transparent occupation-side semantic proxies. Because the benchmark datasets do not expose standardized wage and vacancy tables in the same way as product datasets expose price or discount information, the study constructs six interpretable proxies from the item metadata and observed transition structure: market prevalence, skill breadth, digital skill density, innovation intensity, role level, and transition mobility. Table~\ref{tab:criteria_proxies} summarizes these proxies and their operationalization. The exact source fields differ slightly between JobHop and Karrierewege, but the semantic definitions of the proxies are kept aligned across both benchmarks. The criterion matrix is denoted by
\begin{equation}
X \in \mathbb{R}^{|\mathcal{I}| \times 6}.
\end{equation}
Global criterion weights are derived from entropy weighting, while user-conditioned weights are obtained from the recency-weighted criterion profile of the user's training history. The final weight vector used by TOPSIS is given in Eq.~\eqref{eq:topsismix}:
\begin{equation}
w_u = \alpha\, w^{(u)} + (1-\alpha)\, w^{(g)},
\label{eq:topsismix}
\end{equation}
where $w^{(u)}$ denotes the user-conditioned criterion vector, $w^{(g)}$ denotes the global entropy-based criterion vector, and $\alpha \in [0,1]$ is validation-selected. Standard TOPSIS is then applied to produce a normalized closeness score $s^{\mathrm{T}}_{u,i}$ for each candidate occupation. This choice is motivated by the interpretability and broad applicability of TOPSIS-based multi-criteria ranking in recommender and decision-support settings \citep{zavadskas2016development, monti2021systematic, baczkiewicz2021mcdm, mukhametzyanov2021specific}.

For completeness, if $D^{+}_{u,i}$ and $D^{-}_{u,i}$ denote the weighted Euclidean distances of occupation $i$ from the ideal and anti-ideal points under user-specific weights, then the final TOPSIS closeness score in Eq.~\eqref{eq:topsis} is
\begin{equation}
s^{\mathrm{T}}_{u,i} = \frac{D^{-}_{u,i}}{D^{+}_{u,i} + D^{-}_{u,i}}.
\label{eq:topsis}
\end{equation}

\begin{table}[pos=!htbp]
	\centering
	\caption{Semantic proxies used in the TOPSIS branch.}
	\label{tab:criteria_proxies}
	\small
	\renewcommand{\arraystretch}{1.3}
	\begin{tabularx}{\linewidth}{@{} l >{\raggedright\arraybackslash}X @{}}
		\toprule
		\textbf{Proxy} & \textbf{Operationalization} \\
		\midrule
		Market prevalence & Normalized occupation frequency in the training histories. \\
		Skill breadth & Number of extracted skills or informative description terms associated with the occupation. \\
		Digital skill density & Share of occupation-side terms matching a curated digital-skills lexicon. \\
		Innovation intensity & Presence of innovation-oriented terms such as AI, cloud, security, automation, IoT, or blockchain. \\
		Role level & Heuristic score from title cues such as technician, analyst, engineer, architect, or manager. \\
		Transition mobility & Number of distinct outgoing occupation transitions observed in the training data. \\
		\bottomrule
	\end{tabularx}
\end{table}

The six proxies in Table~\ref{tab:criteria_proxies} jointly span market structure (prevalence, transition mobility), occupational depth (skill breadth, digital skill density), thematic novelty (innovation intensity), and seniority (role level), giving the TOPSIS branch a compact but multidimensional view of each candidate occupation without requiring external wage or vacancy tables.

\medskip
\noindent\textit{Ablation design.}
The proposed full hybrid is evaluated alongside single-branch and pairwise competitors. The baseline set includes popularity, repeat-last, item Markov, transition-aware CF, non-negative matrix factorization (NMF), truncated singular value decomposition (SVD), GRU4Rec, SASRec, TOPSIS-only, RL-only, CF+TOPSIS, and RL+TOPSIS\@. GRU4Rec and SASRec are included as strong neural sequential references because the task is inherently chronological and because recurrent and self-attentive recommenders remain important comparator families for sequence-aware ranking problems \citep{hidasi2015gru4rec, kang2018sasrec}. This ablation structure is essential because the central claim of the paper is not that ``hybridization helps'' in the abstract, but that a specific interaction among collaborative, reinforcement-style, and multi-criteria signals yields measurable benefit under fixed chronological evaluation. Figure~\ref{fig:talent_method_pipeline} makes the scoring logic explicit: all branches operate on the same frozen benchmark inputs, branch scores remain visible until the late-fusion stage, and validation-selected weights are applied only after the separate branch outputs have been computed.

\begin{plainfigurefloat}[!htbp]
\begin{minipage}{\textwidth}
	\centering
	\includegraphics[width=1\textwidth]{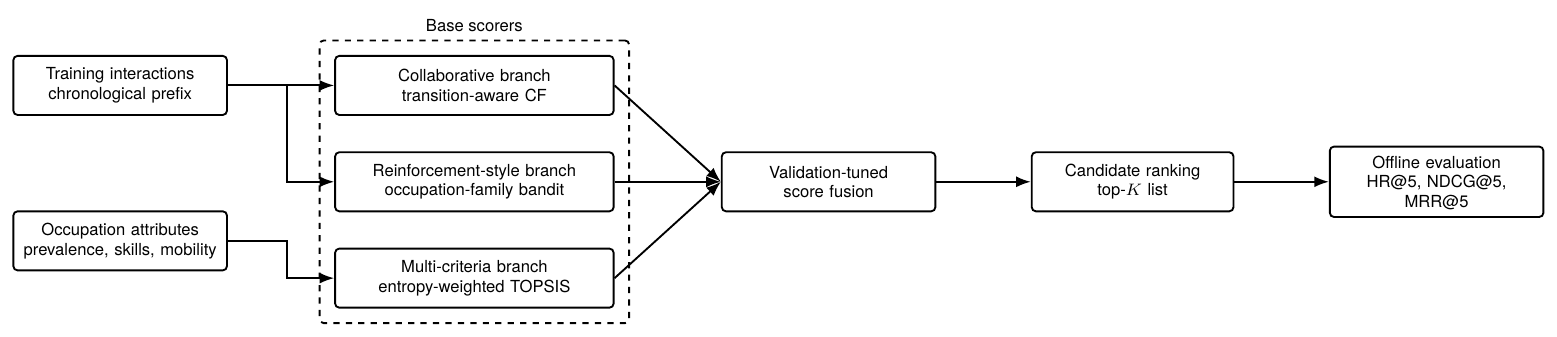}
	\captionof{figure}{CF-RL-TOPSIS late-fusion scoring pipeline.}
	\label{fig:talent_method_pipeline}
\end{minipage}
\end{plainfigurefloat}

The arrows in Figure~\ref{fig:talent_method_pipeline} make two design properties visible at a glance: the three branches consume the same training prefix in parallel, and the late-fusion node only operates on the resulting score vectors. This separation is what allows each branch to be tuned, ablated, or swapped without disturbing the others, and it is also what keeps the validation-selected coefficients $\lambda_{\mathrm{CF}}$, $\lambda_{\mathrm{RL}}$, $\lambda_{\mathrm{T}}$ readable as direct evidence of branch-level contribution rather than as opaque hyperparameters.

\subsection{Evaluation Protocol}

All models are evaluated with top-$5$ ranking metrics, and the manuscript tables report hit ratio (HR@5), normalized discounted cumulative gain (NDCG@5), and mean reciprocal rank (MRR@5). Precision@5 is also tracked internally, but under a single-relevant-item setup it is only a fixed rescaling of HR@5 and is therefore omitted from the main tables. This choice follows directly from the task definition. Next-occupation recommendation is a ranked retrieval problem rather than a binary classification problem: the system must not only recover the correct held-out occupation, but also place it as high as possible in the candidate list.

Because each user contributes one relevant test occupation per split, NDCG@5 and MRR@5 are especially informative: they indicate not only whether the correct occupation appears in the top-$5$, but also how high it is placed. In contrast, a set-based classification view would suppress that ranking information and would therefore be less well aligned with the actual recommendation objective. Rankings are computed over the full retained occupation universe rather than after removing previously observed occupations, because remaining in or returning to a previously held occupation is a valid career outcome in historical talent trajectories.

Model selection is performed on the validation split within each chronological split. For TOPSIS, the mixing coefficient $\alpha$ is selected from $\{0.0, 0.25, 0.5, 0.75, 1.0\}$. For the pairwise and full hybrids, the collaborative weight is searched over $\{0.0, 0.1, \dots, 0.7\}$ and the reinforcement-style weight over $\{0.0, 0.1, \dots, 0.5\}$, with the remaining mass assigned to TOPSIS\@. Matrix-factorization baselines are validation-tuned over latent dimensions $\{6, 10, 14\}$.

GRU4Rec uses a single embedding layer, one GRU layer, and a linear output layer over the retained occupation vocabulary, while SASRec uses item and positional embeddings, one self-attention block, two heads, a feed-forward sublayer of width $2d$, dropout 0.1, and a linear output layer over the same vocabulary. Under the hidden-dimension grid and benchmark vocabularies used here, these neural baselines span approximately 2.8k--31.1k trainable parameters for GRU4Rec and 3.7k--40.2k for SASRec. The RL-bandit settings are otherwise fixed by design across all experiments (learning rate $0.2$, negative reward $-0.2$, two sampled negatives per positive transition, 30 training passes, transition weight $0.7$), while the neural baselines use learning rate $10^{-3}$, weight decay $10^{-5}$, and batch size 64. The 25-epoch training budget for GRU4Rec and SASRec is calibrated to the small-vocabulary, short-sequence regime of these ICT benchmarks (47 and 35 retained occupations; mean sequence length below five), in which longer training schedules are not expected to materially change the validation-tuned hidden dimension.

As a targeted robustness check on whether a deeper RL approximation would alter the main conclusion, the study also compares the tabular family bandit with a compact family-level deep Q-network (DQN) tuned over hidden sizes $\{16, 32, 64\}$ under the same validation protocol. The DQN uses a concatenated current-family and normalized family-frequency state vector, two ReLU-activated hidden layers, and 582--5382 trainable parameters across the tuned hidden sizes.

The repeated-split analysis reports mean and population standard deviation over 10 chronological splits and uses paired Wilcoxon signed-rank comparisons for the full hybrid against a small set of pre-specified strong competitors, together with rank-biserial effect sizes as companion magnitude estimates \citep{wilcoxon1945individual, cureton1956rankbiserial}. Because these comparisons are planned rather than post hoc, raw $p$-values are reported without an additional multiplicity correction and are interpreted jointly with the accompanying effect sizes at a nominal 0.05 significance level. To contextualize absolute deltas, paired Cohen's $d_z$ values are also inspected for the strongest NDCG@5 comparisons. This keeps the significance analysis focused on the principal empirical question rather than on a large post hoc family of pairwise comparisons.

The evaluation is strictly offline and should be interpreted accordingly. Following the cautionary recommendations in the offline-recommendation literature \citep{zhao2022offlineeval, jadidinejad2021simpson, carraro2021debiasing}, the paper does not treat metric values as portable across benchmarks or directly comparable to values reported for different recommendation tasks. Instead, it focuses on controlled relative comparisons under a stable protocol and on how benchmark regime differences alter the usefulness of the individual branches. This distinction is especially important here because a more realistic chronological ranking setup can yield lower absolute metric values than simpler or more weakly constrained evaluation settings while still providing stronger scientific evidence.

\section{Results}\label{sec:results}

The results are presented in the same order as the central argument: JobHop first, Karrierewege second, then the cross-benchmark interpretation and the supporting robustness checks.

\subsection{Primary Results on JobHop}

JobHop serves as the primary benchmark because it yields the clearest evidence for the proposed method. Table~\ref{tab:main_results} reports the repeated-split ranking results for all baselines and hybrid variants. Under repeated chronological splits, the full hybrid achieves the strongest overall ranking quality with $\mathrm{HR@5}=0.4213 \pm 0.0093$, $\mathrm{NDCG@5}=0.3040 \pm 0.0073$, and $\mathrm{MRR@5}=0.2656 \pm 0.0073$. The most competitive classical non-hybrid baselines are item Markov with $\mathrm{NDCG@5}=0.2937 \pm 0.0051$ and transition-aware collaborative filtering with $\mathrm{NDCG@5}=0.2932 \pm 0.0066$. Among the neural sequential references, GRU4Rec reaches $\mathrm{NDCG@5}=0.2835 \pm 0.0054$ and SASRec reaches $\mathrm{NDCG@5}=0.2945 \pm 0.0065$. The strongest pairwise hybrid, CF+TOPSIS, reaches $\mathrm{NDCG@5}=0.2924 \pm 0.0057$.

\begin{table}[pos=htbp]
	\centering
	\caption{Repeated-split ranking results over 10 chronological splits.}
	\label{tab:main_results}
	\renewcommand{\arraystretch}{1.1}
	\resizebox{\linewidth}{!}{%
		\begin{tabular}{lcccccc}
			\toprule
			\multirow{2}{*}{\textbf{Model}} & \multicolumn{3}{c}{\textbf{JobHop}} & \multicolumn{3}{c}{\textbf{Karrierewege}} \\
			\cmidrule(lr){2-4}\cmidrule(lr){5-7}
			& \textbf{HR@5} & \textbf{NDCG@5} & \textbf{MRR@5} & \textbf{HR@5} & \textbf{NDCG@5} & \textbf{MRR@5} \\
			\midrule
			Popularity & 0.2889 $\pm$ 0.0025 & 0.2026 $\pm$ 0.0022 & 0.1745 $\pm$ 0.0024 & 0.5180 $\pm$ 0.0064 & 0.3817 $\pm$ 0.0036 & 0.3374 $\pm$ 0.0031 \\
			Repeat-last & 0.3626 $\pm$ 0.0039 & 0.2819 $\pm$ 0.0034 & 0.2551 $\pm$ 0.0034 & 0.7047 $\pm$ 0.0043 & 0.6034 $\pm$ 0.0041 & 0.5695 $\pm$ 0.0041 \\
			Item Markov & 0.3986 $\pm$ 0.0062 & 0.2937 $\pm$ 0.0051 & 0.2592 $\pm$ 0.0049 & \textbf{0.7506 $\pm$ 0.0058} & 0.6218 $\pm$ 0.0056 & 0.5790 $\pm$ 0.0058 \\
			Transition-CF & 0.4024 $\pm$ 0.0059 & 0.2932 $\pm$ 0.0066 & 0.2573 $\pm$ 0.0069 & 0.7332 $\pm$ 0.0119 & 0.6156 $\pm$ 0.0086 & 0.5764 $\pm$ 0.0081 \\
			NMF & 0.3404 $\pm$ 0.0076 & 0.2461 $\pm$ 0.0043 & 0.2154 $\pm$ 0.0037 & 0.5275 $\pm$ 0.0098 & 0.4743 $\pm$ 0.0057 & 0.4568 $\pm$ 0.0047 \\
			SVD & 0.3352 $\pm$ 0.0116 & 0.2427 $\pm$ 0.0060 & 0.2125 $\pm$ 0.0045 & 0.5277 $\pm$ 0.0128 & 0.4739 $\pm$ 0.0068 & 0.4563 $\pm$ 0.0052 \\
			GRU4Rec & 0.3907 $\pm$ 0.0069 & 0.2835 $\pm$ 0.0054 & 0.2483 $\pm$ 0.0055 & 0.6607 $\pm$ 0.0151 & 0.5632 $\pm$ 0.0147 & 0.5309 $\pm$ 0.0150 \\
			SASRec & 0.4093 $\pm$ 0.0078 & 0.2945 $\pm$ 0.0065 & 0.2568 $\pm$ 0.0066 & 0.7327 $\pm$ 0.0099 & 0.6125 $\pm$ 0.0071 & 0.5724 $\pm$ 0.0066 \\
			TOPSIS & 0.3173 $\pm$ 0.0048 & 0.2117 $\pm$ 0.0029 & 0.1773 $\pm$ 0.0027 & 0.5309 $\pm$ 0.0107 & 0.3825 $\pm$ 0.0064 & 0.3337 $\pm$ 0.0061 \\
			RL-bandit & 0.3835 $\pm$ 0.0087 & 0.2726 $\pm$ 0.0065 & 0.2362 $\pm$ 0.0067 & 0.6224 $\pm$ 0.0044 & 0.4755 $\pm$ 0.0034 & 0.4262 $\pm$ 0.0032 \\
			CF+TOPSIS & 0.3997 $\pm$ 0.0082 & 0.2924 $\pm$ 0.0057 & 0.2573 $\pm$ 0.0060 & 0.7449 $\pm$ 0.0066 & 0.6208 $\pm$ 0.0059 & 0.5797 $\pm$ 0.0058 \\
			RL+TOPSIS & 0.3857 $\pm$ 0.0096 & 0.2629 $\pm$ 0.0087 & 0.2224 $\pm$ 0.0090 & 0.6700 $\pm$ 0.0057 & 0.5064 $\pm$ 0.0110 & 0.4517 $\pm$ 0.0139 \\
			\textbf{CF+RL+TOPSIS} & \textbf{0.4213 $\pm$ 0.0093} & \textbf{0.3040 $\pm$ 0.0073} & \textbf{0.2656 $\pm$ 0.0073} & \textbf{0.7458 $\pm$ 0.0051} & \textbf{0.6227 $\pm$ 0.0053} & \textbf{0.5818 $\pm$ 0.0055} \\
			\bottomrule
		\end{tabular}%
	}
\end{table}

The proposed full hybrid outperforms repeat-last, item Markov, transition-aware CF, CF+TOPSIS, GRU4Rec, and SASRec. The paired Wilcoxon signed-rank tests yield $p=0.002$ in the repeat-last, item-Markov, transition-CF, CF+TOPSIS, and GRU4Rec comparisons, while the full-hybrid-versus-SASRec comparison remains significant at $p=0.0039$ with $r_{rb}=0.9636$. The remaining JobHop effect sizes are maximal ($r_{rb}=1.0$), indicating near-complete split-wise directional consistency in favor of the full hybrid.

Relative to item Markov, the gain is approximately 3.5\% in NDCG@5; relative to CF+TOPSIS, the gain is approximately 4.0\%. Although these raw margins are modest, the paired standardized effects remain large on JobHop: Cohen's $d_z$ is approximately $1.87$ versus item Markov, $2.77$ versus CF+TOPSIS, and $1.20$ versus SASRec. These large standardized values reflect the very low across-split variance produced by fixed benchmark seeds rather than large absolute differences, and they should be read as evidence of consistent directional dominance under the protocol rather than as a measure of deployment-scale practical impact.

The JobHop ablations also identify which partial combinations remain insufficient. RL+TOPSIS stays below both the collaborative baselines and CF+TOPSIS, while TOPSIS-only is substantially weaker still. This indicates that the collaborative branch is the structural backbone of the method, but it also shows that the best collaborative solution is not CF alone. The measurable gap between CF+TOPSIS and the full hybrid implies that the reinforcement-style branch adds measurable value on this benchmark when attached to a strong collaborative core rather than used as a standalone replacement for it. Figure~\ref{fig:selected_models} makes the benchmark contrast visible: on JobHop the full hybrid separates from both the strongest sequential baselines and the pairwise hybrids, whereas on Karrierewege the gaps between models are visibly more compressed.

\begin{plainfigurefloat}[!htbp]
\begin{minipage}{\textwidth}
	\centering
	\includegraphics[width=0.90\textwidth]{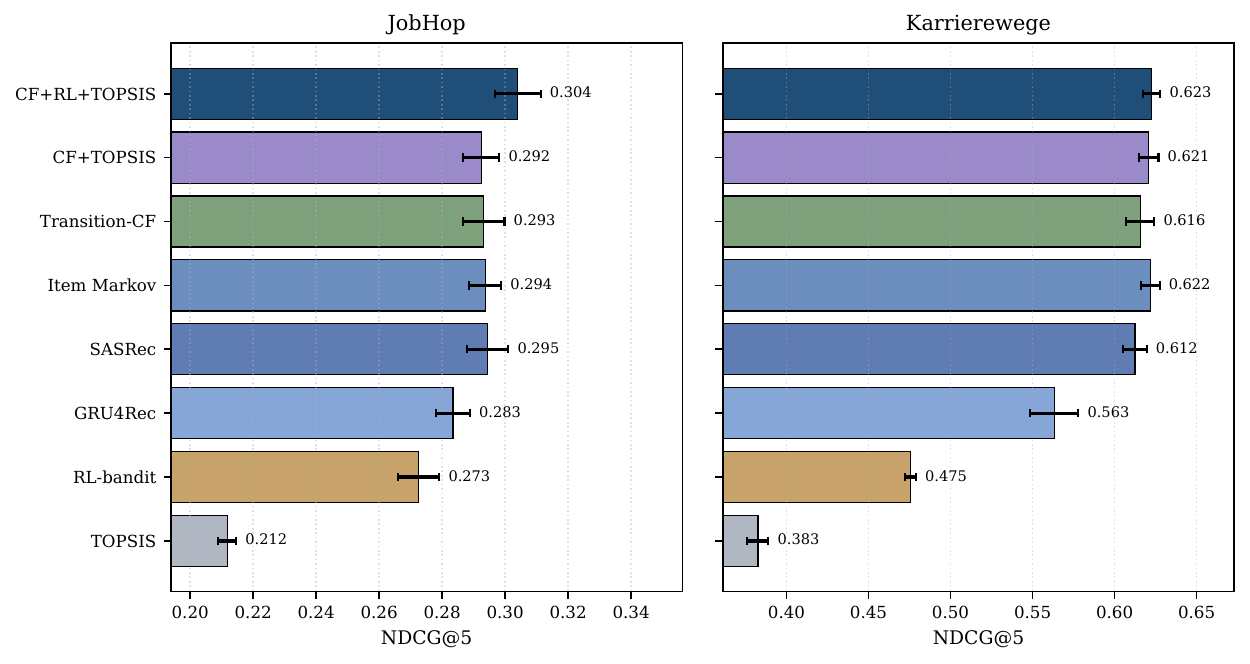}
	\captionof{figure}{Repeated-split NDCG@5 for selected baselines and hybrids on the two ICT benchmarks.}
	\label{fig:selected_models}
\end{minipage}
\end{plainfigurefloat}

The visual gap on the JobHop side of Figure~\ref{fig:selected_models} matches the significant Wilcoxon comparisons reported above, whereas the visibly compressed Karrierewege side anticipates the more cautious reading that the secondary benchmark requires.

\subsection{Secondary Results on Karrierewege}

Karrierewege serves a different role. On this benchmark, the full hybrid remains competitive and reaches $\mathrm{HR@5}=0.7458 \pm 0.0051$, $\mathrm{NDCG@5}=0.6227 \pm 0.0053$, and $\mathrm{MRR@5}=0.5818 \pm 0.0055$. However, the strongest sequential baseline, item Markov, already attains $\mathrm{NDCG@5}=0.6218 \pm 0.0056$, and the difference between the two is not statistically significant ($p=0.5566$, $r_{rb}=0.2364$). Among the neural sequence models, GRU4Rec reaches $\mathrm{NDCG@5}=0.5632 \pm 0.0147$ and SASRec reaches $\mathrm{NDCG@5}=0.6125 \pm 0.0071$. The full hybrid therefore still exceeds repeat-last, transition-aware CF, GRU4Rec, and SASRec, but it no longer creates the clear separation over item Markov observed on JobHop.

This outcome strengthens rather than weakens the role of Karrierewege as a robustness benchmark: it shows that the proposed model is not a universal replacement for simple sequential heuristics, and that its gains depend on whether the benchmark leaves enough room for adaptive and multi-criteria effects beyond role persistence.

In practical terms, the Karrierewege margin between the full hybrid and item Markov is only 0.0009 in NDCG@5, while the margin over CF+TOPSIS is 0.0019. The standardized interpretation is consistent with that reading: the full-hybrid-versus-item-Markov difference corresponds to only $d_z \approx 0.22$, whereas the model remains meaningfully ahead of GRU4Rec and SASRec. Table~\ref{tab:main_results} reports repeated-split means with population standard deviations, and bold marks the best value within each benchmark-metric column. These gaps are too small to sustain a strong superiority claim over item Markov on this benchmark.

\subsection{Cross-Benchmark Performance Differences}

The cross-benchmark contrast is visible not only in the final ranking metrics but also in the planned paired comparisons. Table~\ref{tab:weights_and_tests} collects the Wilcoxon and rank-biserial results for the full hybrid against the strongest pre-specified baselines. For clarity, rank-biserial values of $r_{rb}=1.0000$ indicate complete directional dominance across the non-zero split-wise differences in favor of the full hybrid; they should not be interpreted as perfect predictive accuracy.

\begin{table}[pos=!ht]
	\centering
	\caption{Planned paired comparison statistics for the full hybrid.}
	\label{tab:weights_and_tests}
	\small
	\renewcommand{\arraystretch}{1.2}
	\begin{tabular}{lcc}
		\toprule
		\textbf{Statistic} & \textbf{JobHop} & \textbf{Karrierewege} \\
		\midrule
		Full vs Repeat-last ($p/r_{rb}$) & 0.0020 / 1.0000 & 0.0020 / 1.0000 \\
		Full vs Item Markov ($p/r_{rb}$) & 0.0020 / 1.0000 & 0.5566 / 0.2364 \\
		Full vs Transition-CF ($p/r_{rb}$) & 0.0020 / 1.0000 & 0.0098 / 0.8909 \\
		Full vs CF+TOPSIS ($p/r_{rb}$) & 0.0020 / 1.0000 & 0.0625 / 1.0000 \\
		Full vs GRU4Rec ($p/r_{rb}$) & 0.0020 / 1.0000 & 0.0020 / 1.0000 \\
		Full vs SASRec ($p/r_{rb}$) & 0.0039 / 0.9636 & 0.0020 / 1.0000 \\
		\bottomrule
	\end{tabular}

	\smallskip
	\footnotesize \textit{Note.} $p$: paired Wilcoxon signed-rank $p$-value over 10 chronological splits; $r_{rb}$: rank-biserial correlation as a directional dominance measure for the full hybrid against the listed baseline.
\end{table}

Read together, the Wilcoxon and rank-biserial entries in Table~\ref{tab:weights_and_tests} expose the cross-benchmark asymmetry directly: on JobHop every planned comparison is significant and almost every effect size is maximal, whereas on Karrierewege the comparison against item Markov shrinks to a small, non-significant effect while the comparisons against repeat-last, GRU4Rec, and SASRec remain strong.

The fitted mixture points the same way. Figure~\ref{fig:fusion_weights} visualizes the mean validation-selected branch weights. On JobHop, the average repeated-split full-hybrid weights are approximately $\lambda_{\mathrm{CF}}=0.53$, $\lambda_{\mathrm{RL}}=0.28$, and $\lambda_{\mathrm{T}}=0.19$. On Karrierewege, the corresponding averages are $\lambda_{\mathrm{CF}}=0.69$, $\lambda_{\mathrm{RL}}=0.05$, and $\lambda_{\mathrm{T}}=0.26$. The RL branch therefore remains meaningfully active on JobHop but largely collapses on Karrierewege, where the fusion is pushed almost entirely toward the collaborative and multi-criteria branches.

\begin{plainfigurefloat}[!htbp]
\begin{minipage}{\textwidth}
	\centering
	\includegraphics[width=0.78\textwidth]{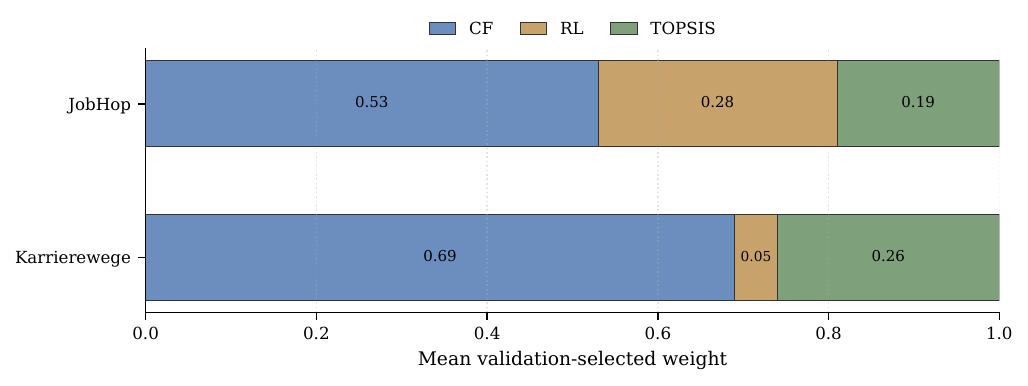}
	\captionof{figure}{Mean validation-selected full-hybrid branch weights on each benchmark.}
	\label{fig:fusion_weights}
\end{minipage}
\end{plainfigurefloat}

The difference is most plausibly explained by benchmark regime. Karrierewege appears to be much more persistence-dominated: repeat-last already reaches $\mathrm{NDCG@5}=0.6034$, and item Markov reaches $0.6218$, leaving only a very narrow margin for richer modeling. In such a regime, the marginal value of an additional reinforcement-style family-transition bias becomes negligible. JobHop, by contrast, is less saturating under persistence baselines: repeat-last remains at $0.2819$ and item Markov at $0.2937$, leaving more room for adaptive and semantic signals beyond simple role repetition. There, the hybrid benefits from a measurable three-way complementarity rather than relying almost entirely on the collaborative branch.

The effect-size profile points the same way: the classical JobHop comparisons all show complete directional dominance of the full hybrid ($r_{rb}=1.0$), and even the stronger self-attentive comparator SASRec remains below the hybrid with near-complete directional dominance ($r_{rb}=0.9636$), whereas on Karrierewege only the item-Markov comparison shrinks to a small effect.

This difference is also consistent with recent calls for more realistic and scenario-sensitive evaluation in career-path prediction \citep{senger2025realistic}. The point is not that one benchmark is ``correct'' and the other is not. Rather, the two datasets expose different recommendation regimes. JobHop leaves enough headroom for collaborative, adaptive, and semantic signals to interact, whereas Karrierewege is closer to a high-persistence continuation task. That distinction is scientifically useful because it helps separate benchmark-sensitive signal interaction from benchmark-independent claims about model quality.

\subsection{Worked JobHop Interpretability Example}

To make the interpretability claim more concrete, Table~\ref{tab:interpretability_case} reports one anonymized canonical-split JobHop user whose training history is \emph{ICT help desk agent $\rightarrow$ web developer $\rightarrow$ web developer $\rightarrow$ ICT application developer $\rightarrow$ web developer}. The held-out next occupation is \emph{user interface developer}. For this user, canonical validation selects $\lambda_{\mathrm{CF}}=0.5$, $\lambda_{\mathrm{RL}}=0.4$, and $\lambda_{\mathrm{T}}=0.1$. For readability, all score columns in the table, including the fused Full column, are min-max normalized within user, and the held-out target is marked with an asterisk. The user-conditioned TOPSIS weights are concentrated mainly on market prevalence (0.32), transition mobility (0.31), and digital skill density (0.18), with the remaining criteria contributing smaller shares.

\begin{table}[pos=!htbp]
	\centering
	\caption{Worked JobHop interpretability example.}
	\label{tab:interpretability_case}
	\small
	\renewcommand{\arraystretch}{1.2}
	\begin{tabular}{lcccc}
		\toprule
		\textbf{Candidate occupation} & \textbf{CF} & \textbf{RL} & \textbf{TOPSIS} & \textbf{Full} \\
		\midrule
		Web developer & 1.000 & 1.000 & 0.440 & 1.000 \\
		Software developer & 0.285 & 0.634 & 0.723 & 0.489 \\
		User interface developer* & 0.035 & 0.895 & 0.070 & 0.397 \\
		ICT help desk agent & 0.170 & 0.446 & 1.000 & 0.376 \\
		ICT application developer & 0.249 & 0.508 & 0.302 & 0.370 \\
		\bottomrule
	\end{tabular}

	\smallskip
	\footnotesize $^{*}$Held-out target occupation; recovered at rank 3 in the full hybrid despite low CF and TOPSIS scores, because the family-level RL branch sharply elevates same-family candidates.
\end{table}

This case is informative because the held-out target is ranked third by the full hybrid, but only 21st by transition-aware CF, 22nd by item Markov, 31st by CF+TOPSIS, and 40th by TOPSIS alone. The table shows why. The collaborative branch strongly prefers staying with \emph{web developer}, which is also the top full-hybrid candidate. The RL branch, however, sharply elevates \emph{user interface developer} because it remains within the same digital-experience family as the user's recent role pattern, and that family-level preference is still visible even after popularity smoothing.

TOPSIS contributes only a modest correction in this example because the target occupation is not especially strong on the user-weighted prevalence and mobility criteria. The case therefore illustrates the practical meaning of structural interpretability in the proposed model: branch scores, fusion weights, and criterion emphasis can all be inspected directly rather than inferred after the fact.

The worked example therefore complements the aggregate results with a concrete ranking mechanism. It shows that the hybrid can rescue a plausible adjacent-role target that remains heavily under-ranked by the collaborative and pairwise alternatives, while still keeping every branch contribution directly inspectable at the user level.

\subsection{Proxy Sensitivity, Runtime, and DQN Check}

Because the TOPSIS branch is built from six semantic proxies, a leave-one-proxy-out sensitivity analysis was also run on the full hybrid. Table~\ref{tab:proxy_sensitivity} reports the repeated-split NDCG@5 means obtained after removing one proxy at a time, with deltas measured against the all-proxy full hybrid (0.3040 on JobHop and 0.6227 on Karrierewege). On JobHop, the resulting changes are all smaller than 0.001 in absolute NDCG@5, with the largest decrease appearing when role level is removed. This suggests that no single semantic criterion alone explains the JobHop gain; instead, the TOPSIS branch acts as a modest distributed correction layered on top of the collaborative and bandit signals. On Karrierewege, market prevalence is clearly the most consequential proxy, while the others produce only minor changes. This pattern is consistent with the earlier finding that the secondary benchmark is more persistence- and prevalence-dominated overall.

\begin{table}[pos=!htbp]
	\centering
	\caption{Leave-one-proxy-out sensitivity of the full hybrid.}
	\label{tab:proxy_sensitivity}
	\small
	\renewcommand{\arraystretch}{1.2}
	\begin{tabular}{lll}
		\toprule
		\textbf{Removed proxy} & \textbf{JobHop} & \textbf{Karrierewege} \\
		\midrule
		Market prevalence & 0.3038 (-0.0002) & 0.6140 (-0.0087) \\
		Skill breadth & 0.3043 (+0.0002) & 0.6235 (+0.0009) \\
		Digital skill density & 0.3034 (-0.0007) & 0.6224 (-0.0002) \\
		Innovation intensity & 0.3045 (+0.0005) & 0.6225 (-0.0001) \\
		Role level & 0.3031 (-0.0009) & 0.6218 (-0.0009) \\
		Transition mobility & 0.3038 (-0.0002) & 0.6211 (-0.0016) \\
		\bottomrule
	\end{tabular}
\end{table}

With proxy sensitivity established as a diffuse small-magnitude effect on JobHop and a single-proxy (market-prevalence) effect on Karrierewege, the next robustness check moves to runtime as a way to contextualize the lightweight design choice. Coarse CPU wall-clock timings were collected on a Windows 11 desktop with an Intel Core i5-12500 CPU, 16\,GB RAM, Python 3.12, PyTorch 2.10, NumPy 2.4, and scikit-learn 1.8 over three canonical runs. Full-hybrid construction and validation selection averaged 10.25\,s on JobHop and 7.85\,s on Karrierewege, compared with 11.07\,s and 11.58\,s for GRU4Rec and 20.24\,s and 18.88\,s for SASRec. Test-set scoring itself remained at approximately 0.02\,s for all three model families. These implementation-dependent timings should not be read as hardware-independent complexity bounds, but they do support the practical appeal of keeping the adaptive branch compact rather than replacing the entire recommender with a deeper sequential model.

\begin{table}[pos=!htbp]
	\centering
	\caption{Family-level DQN robustness check on NDCG@5.}
	\label{tab:dqn_check}
	\small
	\renewcommand{\arraystretch}{1.2}
	\begin{tabular}{lcc}
		\toprule
		\textbf{Model} & \textbf{JobHop} & \textbf{Karrierewege} \\
		\midrule
		RL-bandit & 0.2726 $\pm$ 0.0065 & 0.4755 $\pm$ 0.0034 \\
		Family-DQN & 0.2766 $\pm$ 0.0037 & 0.4783 $\pm$ 0.0042 \\
		CF+RL+TOPSIS & 0.3040 $\pm$ 0.0073 & 0.6227 $\pm$ 0.0053 \\
		CF+DQN+TOPSIS & 0.3040 $\pm$ 0.0074 & 0.6238 $\pm$ 0.0056 \\
		\bottomrule
	\end{tabular}
\end{table}

Replacing the tabular family bandit with a small family-level DQN, summarized in Table~\ref{tab:dqn_check} as mean NDCG@5 $\pm$ population standard deviation over 10 chronological splits, slightly improves the adaptive branch in isolation on both benchmarks: JobHop increases from $0.2726$ to $0.2766$ in NDCG@5 ($p=0.0098$), and Karrierewege increases from $0.4755$ to $0.4783$ ($p=0.0371$). However, the full hybrid remains effectively unchanged. On JobHop, the DQN-augmented hybrid reaches $0.3040$ versus $0.3040$ for the original full hybrid ($p=0.9219$). On Karrierewege, the DQN-augmented hybrid reaches $0.6238$ versus $0.6227$ ($p=0.0625$). This means that a deeper family-level RL approximation can refine the adaptive branch itself without materially changing the paper's main conclusion, which in turn supports the practical sufficiency of the lighter bandit branch under the present frozen benchmark regime.

\section{Discussion}\label{sec:discussion}

The results matter less as a single leaderboard outcome than as evidence about benchmark regime, branch complementarity, and the practical meaning of interpretability in talent recommendation. The discussion therefore moves from the central empirical claim to its practical implications, originality, and limits.

\subsection{Interpretation of the Empirical Claim}

The central empirical claim of the paper is benchmark-conditional rather than universal. The collaborative-reinforcement multi-criteria score-fusion architecture yields its clearest gains on a semantically rich public talent-history benchmark where simple persistence is insufficient, while remaining competitive on a second public benchmark whose dynamics are more strongly governed by persistence. This narrower claim is supported by transparent ablation logic and by two public benchmarks with materially different empirical regimes.

A convincing empirical study should report not only where a model improves, but also the conditions under which the effect weakens, and the present results do exactly that. When the benchmark leaves room for adaptive structure, the RL branch receives a meaningful validation-selected weight and helps the full hybrid surpass strong collaborative baselines; when the benchmark is dominated by last-role regularity, the same branch appropriately collapses toward zero. A targeted family-level DQN check points in the same direction: modest branch-level gains do not translate into a materially stronger full hybrid. This ablation-visible behavior is precisely what a late-fusion design is intended to reveal \citep{burke2002hybrid, caromartinez2021explainable}, and it indicates that the RL branch is not incidental even though its absolute share of the fusion weight varies by benchmark.

The paper therefore contributes a benchmark-sensitive empirical result rather than a one-number leaderboard claim. This framing is important in talent recommendation because public datasets remain scarce, heterogeneous, and methodologically consequential \citep{bied2023toward, poncio2024systematicjob, senger2025realistic}. A model that wins clearly on JobHop but only matches the strongest simple baseline on Karrierewege is more informative about the interaction between benchmark dynamics and model design than a study evaluated on only one public dataset.

\subsection{Interpretation of the Primary-Benchmark Gain}

The JobHop result is meaningful on three converging grounds. The full hybrid is not merely better than weak baselines, since it also significantly exceeds item Markov, transition-aware CF, CF+TOPSIS, and both neural sequential references, GRU4Rec and SASRec. At the same time, the RL branch retains a non-trivial share of the final fusion weight on this benchmark, which means the adaptive branch is materially contributing rather than being carried entirely by the collaborative backbone. The benchmark itself is also public, talent-domain aligned, and semantically rich enough that a multi-criteria branch remains interpretable. JobHop therefore becomes more than a favorable dataset: it functions as a testbed on which the complementarity among collaborative, adaptive, and criteria-based evidence can actually be observed.

On its own, an NDCG@5 value of approximately 0.304 may not look high without context. However, the offline-evaluation literature strongly cautions against naive cross-paper comparisons of absolute values \citep{zhao2022offlineeval, jadidinejad2021simpson, carraro2021debiasing}. What matters here is not how the value compares with unrelated recommendation tasks or with classification-style metrics such as F1, but that it is achieved under a fixed chronological ranking protocol and significantly improves on the strongest available alternatives in the same environment. In recommendation studies built on historical data, relative comparative gain under a stable protocol is typically more informative than the raw absolute metric value by itself. A stricter chronological ranking setup can therefore produce lower absolute scores than easier or less constrained evaluation regimes while still yielding stronger scientific evidence.

This result also aligns with the growing labor-market recommendation literature that emphasizes richer representations and adaptive reasoning when text quality, sparse histories, or changing user intent make simple heuristics insufficient \citep{bied2023toward, du2024llmjob, wu2024llmgraphjob}. The present analysis does not depend on LLM-based recruitment augmentation, yet the JobHop outcome supports the same broader message: once the benchmark preserves enough semantic variation, relying solely on last-item continuation becomes unnecessarily restrictive.

The added GRU4Rec and SASRec comparisons further reduce the risk that the JobHop finding is an artifact of a purely classical baseline suite, since the full hybrid still remains ahead when both recurrent and self-attentive sequential references are introduced. Stronger contrastive sequential models such as CL4SRec or graph-based recommenders such as LightGCN are not included here, because the limited vocabulary and short retained histories of these ICT benchmarks (47 and 35 retained occupations, mean sequence length below five) restrict the practical headroom for those families, and the GRU4Rec/SASRec pair already covers the recurrent and self-attentive references that dominate public sequential-recommendation reporting.

\subsection{Practical Interpretation for Talent Systems}

The empirical pattern suggests a practical deployment rule. When the career-history environment is dominated by strong role persistence, simple sequential recommenders may already absorb most of the predictable signal and the RL branch should be expected to contribute little. In contrast, when the benchmark reflects more diverse career movement and semantically richer occupation variation, the combined use of collaborative, adaptive, and criteria-based evidence becomes more worthwhile. As a rough operational signal informed by the two benchmarks studied here, when the persistence baseline (e.g., NDCG@5 of repeat-last) approaches the 0.55--0.65 range, the headroom for adaptive and multi-criteria contributions becomes narrow and the hybrid mainly secures a competitive position rather than a clear lead, retaining its interpretability advantage at no ranking cost; when the persistence baseline remains below approximately 0.30, the proposed hybrid recovers measurable headroom over the strongest sequential alternative.

This pattern translates directly into talent-platform deployment criteria. The architecture is therefore especially attractive in decision-support settings such as role exploration, upskilling guidance, internal mobility, and adjacent-role discovery, where a system is expected not only to repeat the most likely next job but also to surface interpretable and plausibly progressive alternatives.

In quantitative terms, the JobHop NDCG@5 gain of approximately 3.5\% over the strongest classical baseline reflects a measurable, though not transformative, improvement in top-$K$ recovery across the test population. In environments where many small recommendation events aggregate into longer-term career-navigation outcomes, even modest per-recommendation gains can compound at scale, and the explicit branch scores and inspectable fusion weights allow human-resources analysts and career advisors to audit each recommendation against the user's own trajectory rather than relying on an opaque ranked list.

This interpretation connects directly to recent explainable and multi-stakeholder job recommendation work. Candidate-facing talent recommenders are increasingly expected to justify why a role is recommended, not merely to output a ranked list \citep{schellingerhout2024explainablejob, tang2025explainablepjr}. The explicit TOPSIS branch is useful here because it makes the occupation-side rationale inspectable. Even when it is not the strongest standalone predictor, it provides a transparent semantic anchor that pure collaborative or neural models often struggle to expose directly.

The reliance on the term \emph{interpretable} rather than a stronger post-hoc explainable artificial intelligence (XAI) claim is also deliberate. In the explainable-recommendation literature, inherently transparent models and post-hoc explanation layers are treated as related but distinct design choices \citep{zhang2020explainable, tang2025explainablepjr}. The present model belongs to the first category: each branch produces an explicit score, the TOPSIS component exposes criterion weights and ideal-versus-anti-ideal distance logic, the validation-selected fusion weights remain observable, and branch ablations show how much each signal family contributes on each benchmark. The architecture therefore argues for structural interpretability at the model-design level rather than for SHAP-, LIME-, or DALEX-style explanations added after the fact to an otherwise opaque recommender.

\subsection{Role of the Secondary Benchmark}

Karrierewege remains valuable even though it does not produce a decisive advantage over item Markov, because its inclusion changes what the empirical evidence can support. Without a secondary benchmark, the study would remain more exposed to concerns that the dataset had been chosen primarily because the hybrid performs clearly better there. With Karrierewege in the picture, the evidence instead shows that the hybrid remains competitive under a substantially different regime and that its branch behavior changes in a theoretically interpretable way.

This reading aligns with recent calls for more realistic and scenario-sensitive evaluation in career-path prediction \citep{senger2025realistic}. Benchmark realism is not only about chronology or public release; it is also about the underlying sequence dynamics that determine how much room remains for hybrid signal interaction. A method that wins when the benchmark is semantically rich but loses all advantage when trajectory persistence dominates therefore provides a more realistic picture of model utility than a method evaluated on a single favorable dataset, and this is the role Karrierewege plays in the present study.

\subsection{Originality Relative to Prior Hybrids}

The present study is best positioned not as a wholly new foundational recommender algorithm, but as a talent-domain fusion architecture in which collaborative, reinforcement-style, and multi-criteria signals are brought together under a public, reproducible, dual-benchmark evaluation design. This framing reflects the existing pairwise neighbors in hybrid recommendation, collaborative multi-criteria recommendation, and reinforcement-enhanced recommendation \citep{burke2002hybrid, albashiri2018topsiscf, peng2024drlcf, sun2021skillrl, sun2025marketaware}.

The resulting contribution sits between model innovation and evaluation innovation. It pairs a benchmark-sensitive empirical question with an interpretable late-fusion architecture, and supports both with an explicit branch-level account of why the same hybrid behaves differently across JobHop and Karrierewege. Relative to recent talent-recommendation work, the originality lies as much in the evidence structure used to study the fusion as in the fusion itself.

\subsection{Limitations}

Two clusters of limitations qualify the empirical evidence reported above. Both benchmarks are historical sequence datasets rather than exposure-controlled online systems, and the ICT subsets are constructed through a documented manual allow-list over English-language job titles, so transfer to other languages or to non-ICT domains would require both the allow-list and the proxy lexicon to be reconstructed. The multi-criteria branch likewise relies on semantic proxies rather than official wage, demand, or labor-market outcome tables; the leave-one-proxy-out results indicate that these proxies operate as a diffuse semantic correction rather than as individually dominant drivers on JobHop, but stronger external validation of the proxy design would still be valuable.

On the architectural side, the reinforcement-style component remains a compact bandit signal layered on top of the collaborative backbone rather than a full online deep-RL recommender, and its occupation-family taxonomy is a fixed design choice that alternative groupings (such as a denser ESCO hierarchy or a clustering induced from the transition graph) could plausibly refine. As an operational boundary, the hybrid's gain is also expected to be smaller for users whose training prefix is at the minimum retained length of three or whose history is dominated by exact role repetition, since in those cases the collaborative core already absorbs most of the predictable signal, a pattern consistent with the cross-benchmark contrast reported in Section~\ref{sec:results}.

\section{Conclusion}\label{sec:conclusion}

The study presented a collaborative-reinforcement multi-criteria score-fusion model for skills-aware technology talent recommendation and evaluated it on two frozen public ICT benchmarks against both classical and neural sequential baselines. The clearest empirical result appears on JobHop, where the full CF+RL+TOPSIS hybrid significantly outperforms repeat-last, item Markov, transition-aware collaborative filtering, CF+TOPSIS, GRU4Rec, and SASRec under repeated chronological evaluation. On Karrierewege, the method remains competitive but does not significantly exceed the strongest Markov baseline, indicating a benchmark regime governed more strongly by persistence and leaving substantially less room for an adaptive branch to help. Taken together, the two-benchmark findings support a benchmark-sensitive rather than benchmark-agnostic claim.

A transparent late-fusion architecture can separate collaborative structure, reinforcement-style adaptation, and occupation-side criteria strongly enough to reveal when these signals complement one another and when one branch should become inactive. In practical terms, the proposed hybrid is best matched to semantically rich, non-saturating transition problems in which simple continuation heuristics do not already absorb most of the available signal, and it remains competitive in persistence-dominated regimes through its strong collaborative backbone while preserving the interpretability advantage of explicit branch scores and fusion weights. The contribution is therefore not a benchmark-agnostic claim of universal superiority, but a reproducible account of why the same hybrid behaves differently under two public talent-history regimes.

Future work can extend the framework along several directions while preserving the present emphasis on frozen benchmarks, repeated chronological reporting, and ablation-visible branch behavior. The most important extension is to combine the present public benchmark core with an additional software-jobs validation layer tied more explicitly to current labor-market skill resources, and to test richer contextual or session-aware RL branches without sacrificing the current ablation discipline. A complementary direction is to construct a Karrierewege subset that excludes users whose training prefix is dominated by exact role repetition, in order to test whether the bandit branch reactivates once the persistence floor is reduced. A further line of work is to add stakeholder-facing explanation studies, since recent person--job recommendation research increasingly treats explainability as a primary evaluation dimension rather than a secondary reporting aid \citep{schellingerhout2024explainablejob, tang2025explainablepjr}. Future evaluations should also continue to report when an added branch materially changes the ranking and when it merely adds complexity, since that benchmark-aware discipline is what gives a hybrid recommender its scientific traction.

\section*{CRediT authorship contribution statement}
\textbf{\"Ozkan Canay:} Conceptualization, Methodology, Software, Validation, Formal analysis, Investigation, Data curation, Visualization, Writing -- original draft, Writing -- review \& editing.

\section*{Declaration of Generative AI and AI-assisted technologies in the writing process}
During the preparation of this work, the author used OpenAI ChatGPT and Anthropic Claude to support English translation, language refinement, academic editing, and manuscript restructuring. The author also used these tools for limited code-level prototyping support. After using these tools, the author reviewed and edited all AI-assisted outputs as needed and takes full responsibility for the content of the publication.

\section*{Declaration of competing interest}
The author declares that there are no known competing financial interests or personal relationships that could have appeared to influence the work reported in this paper.

\section*{Acknowledgment}
This research did not receive any specific grant from funding agencies in the public, commercial, or not-for-profit sectors.

\section*{Data Availability}
The data and code that support the findings of this study are openly available at \url{https://github.com/canay/hybrid-talent-rec-benchmarks}. The original public JobHop and Karrierewege source files can be obtained from the providers linked in the repository. The repository additionally exposes the digital-skills lexicon and role-level cue list used by the TOPSIS proxies in machine-readable form, so that the proxy construction can be inspected and adapted independently. Both source datasets are publicly released and de-identified, so no separate institutional review board approval was required for this secondary use.

\bibliographystyle{cas-model2-names}
\bibliography{cas-refs}

@article{albashiri2018topsiscf,
  author  = {Al-bashiri, Hael and Abdulgabber, Mansoor Abdullateef and Romli, Awanis},
  title   = {An Improved Memory-Based Collaborative Filtering Method Based on the TOPSIS Technique},
  journal = {PLOS ONE},
  year    = {2018},
  volume  = {13},
  number  = {10},
  pages   = {e0204434},
  doi     = {10.1371/journal.pone.0204434}
}

@article{baczkiewicz2021mcdm,
  author  = {B{\k{a}}czkiewicz, Aleksandra and Kizielewicz, Bart{\l}omiej and Shekhovtsov, Andrii},
  title   = {Methodical Aspects of MCDM Based E-Commerce Recommender System},
  journal = {Journal of Theoretical and Applied Electronic Commerce Research},
  year    = {2021},
  volume  = {16},
  number  = {6},
  pages   = {2192--2229},
  doi     = {10.3390/jtaer16060122}
}

@article{peng2024drlcf,
  author  = {Peng, Sony and Siet, Sophort and Ilkhomjon, Sadriddinov},
  title   = {Integration of Deep Reinforcement Learning with Collaborative Filtering for Movie Recommendation Systems},
  journal = {Applied Sciences},
  year    = {2024},
  volume  = {14},
  number  = {3},
  pages   = {1155},
  doi     = {10.3390/app14031155}
}

@article{zhao2022offlineeval,
  author  = {Zhao, Wayne Xin and Lin, Zihan and Feng, Zhichao and Wang, Haifeng and Wen, Ji-Rong and Li, Jian},
  title   = {A Revisiting Study of Appropriate Offline Evaluation for Top-N Recommendation Algorithms},
  journal = {ACM Transactions on Information Systems},
  year    = {2022},
  volume  = {41},
  number  = {2},
  pages   = {1--41},
  doi     = {10.1145/3545796}
}

@article{jadidinejad2021simpson,
  author  = {Jadidinejad, Amir H. and Macdonald, Craig and Ounis, Iadh},
  title   = {The Simpson's Paradox in the Offline Evaluation of Recommendation Systems},
  journal = {ACM Transactions on Information Systems},
  year    = {2021},
  volume  = {40},
  number  = {1},
  pages   = {1--22},
  doi     = {10.1145/3458509}
}

@article{carraro2021debiasing,
  author  = {Carraro, Diego and Bridge, Derek},
  title   = {A Sampling Approach to Debiasing the Offline Evaluation of Recommender Systems},
  journal = {Journal of Intelligent Information Systems},
  year    = {2021},
  volume  = {58},
  number  = {2},
  pages   = {311--336},
  doi     = {10.1007/s10844-021-00651-y}
}

@article{wilcoxon1945individual,
  author  = {Wilcoxon, Frank},
  title   = {Individual Comparisons by Ranking Methods},
  journal = {Biometrics Bulletin},
  year    = {1945},
  volume  = {1},
  number  = {6},
  pages   = {80--83},
  doi     = {10.2307/3001968}
}

@article{chen2020attributeaware,
  author  = {Chen, Wen-Hao and Hsu, Chin-Chi and Lai, Yi-An},
  title   = {Attribute-Aware Recommender System Based on Collaborative Filtering: Survey and Classification},
  journal = {Frontiers in Big Data},
  year    = {2020},
  volume  = {2},
  pages   = {49},
  doi     = {10.3389/fdata.2019.00049}
}

@article{caromartinez2021explainable,
  author  = {Caro-Mart{\'i}nez, Marta and Jim{\'e}nez-D{\'i}az, Guillermo and Recio-Garc{\'i}a, Juan A.},
  title   = {Conceptual Modeling of Explainable Recommender Systems: An Ontological Formalization to Guide Their Design and Development},
  journal = {Journal of Artificial Intelligence Research},
  year    = {2021},
  volume  = {71},
  pages   = {557--589},
  doi     = {10.1613/jair.1.12789}
}

@article{roy2022systematic,
  author  = {Roy, Deepjyoti and Dutta, Mala},
  title   = {A Systematic Review and Research Perspective on Recommender Systems},
  journal = {Journal of Big Data},
  year    = {2022},
  volume  = {9},
  number  = {1},
  pages   = {59},
  doi     = {10.1186/s40537-022-00592-5}
}

@article{li2024recent,
  author  = {Li, Yang and Liu, Kangbo and Satapathy, Ranjan and Wang, Suhang and Cambria, Erik},
  title   = {Recent Developments in Recommender Systems: A Survey},
  journal = {IEEE Computational Intelligence Magazine},
  year    = {2024},
  volume  = {19},
  number  = {2},
  pages   = {78--95},
  doi     = {10.1109/MCI.2024.3363984}
}

@article{alhasan2024survey,
  author  = {Al-Hasan, Tamim M. and Sayed, Aya Nabil and Bensaali, Fay{\c{c}}al and Himeur, Yassine and Varlamis, Iraklis and Dimitrakopoulos, George},
  title   = {From Traditional Recommender Systems to GPT-Based Chatbots: A Survey of Recent Developments and Future Directions},
  journal = {Big Data and Cognitive Computing},
  year    = {2024},
  volume  = {8},
  number  = {4},
  pages   = {36},
  doi     = {10.3390/bdcc8040036}
}

@article{lin2023survey,
  author  = {Lin, Yuanguo and Liu, Yong and Lin, Fan and Zou, Lixiang and Wu, Pengcheng and Zeng, Wen and Miao, Chunyan},
  title   = {A Survey on Reinforcement Learning for Recommender Systems},
  journal = {IEEE Transactions on Neural Networks and Learning Systems},
  year    = {2024},
  volume  = {35},
  number  = {10},
  pages   = {13164--13184},
  doi     = {10.1109/TNNLS.2023.3280161}
}

@article{monti2021systematic,
  author  = {Monti, Diego and Rizzo, Giuseppe and Morisio, Maurizio},
  title   = {A Systematic Literature Review of Multicriteria Recommender Systems},
  journal = {Artificial Intelligence Review},
  year    = {2021},
  volume  = {54},
  number  = {1},
  pages   = {427--468},
  doi     = {10.1007/s10462-020-09851-4}
}

@article{burke2002hybrid,
  author  = {Burke, Robin},
  title   = {Hybrid Recommender Systems: Survey and Experiments},
  journal = {User Modeling and User-Adapted Interaction},
  year    = {2002},
  volume  = {12},
  number  = {4},
  pages   = {331--370},
  doi     = {10.1023/A:1021240730564}
}

@article{abdulhussien2021behavior,
  author  = {Abdul Hussien, Farah Tawfiq and Rahma, Abdul Monem S. and Abdulwahab, Hala Bahjat},
  title   = {An E-Commerce Recommendation System Based on Dynamic Analysis of Customer Behavior},
  journal = {Sustainability},
  year    = {2021},
  volume  = {13},
  number  = {19},
  pages   = {10786},
  doi     = {10.3390/su131910786}
}

@article{zhang2023service,
  author  = {Zhang, En and Ma, Wenming and Zhang, Jinkai and Xia, Xuchen},
  title   = {A Service Recommendation System Based on Dynamic User Groups and Reinforcement Learning},
  journal = {Electronics},
  year    = {2023},
  volume  = {12},
  number  = {24},
  pages   = {5034},
  doi     = {10.3390/electronics12245034}
}

@inproceedings{zhao2021dear,
  author    = {Zhao, Xiangyu and Gu, Changsheng and Zhang, Haoshenglun and Yang, Xiwang and Liu, Xiaobing and Tang, Jiliang and Liu, Hui},
  title     = {DEAR: Deep Reinforcement Learning for Online Advertising Impression in Recommender Systems},
  booktitle = {Proceedings of the AAAI Conference on Artificial Intelligence},
  year      = {2021},
  volume    = {35},
  number    = {1},
  pages     = {750--758},
  doi       = {10.1609/aaai.v35i1.16156}
}

@inproceedings{xin2020selfrl,
  author    = {Xin, Xin and Karatzoglou, Alexandros and Arapakis, Ioannis and Jose, Joemon M.},
  title     = {Self-Supervised Reinforcement Learning for Recommender Systems},
  booktitle = {Proceedings of the 43rd International ACM SIGIR Conference on Research and Development in Information Retrieval},
  year      = {2020},
  pages     = {931--940},
  doi       = {10.1145/3397271.3401147}
}

@article{tiwary2024review,
  author  = {Tiwary, Neeraj and Noah, Shahrul Azman Mohd and Fauzi, Fariza and Yee, Tan Su},
  title   = {A Review of Explainable Recommender Systems Utilizing Knowledge Graphs and Reinforcement Learning},
  journal = {IEEE Access},
  year    = {2024},
  volume  = {12},
  pages   = {91999--92019},
  doi     = {10.1109/ACCESS.2024.3422416}
}

@inproceedings{intayoad2018reinforcement,
  title={Reinforcement learning for online learning recommendation system},
  author={Intayoad, W. and Kamyod, C. and Temdee, P.},
  booktitle={2018 Global Wireless Summit (GWS)},
  pages={167--170},
  year={2018},
  doi={10.1109/GWS.2018.8686513}
}

@article{tang2019reinforcement,
  title={A reinforcement learning approach to personalized learning recommendation systems},
  author={Tang, X. and Chen, Y. and Li, X. and Liu, J. and Ying, Z.},
  journal={British Journal of Mathematical and Statistical Psychology},
  volume={72},
  number={1},
  pages={108--135},
  year={2019},
  doi={10.1111/bmsp.12144}
}

@inproceedings{munemasa2018deep,
  title={Deep reinforcement learning for recommender systems},
  author={Munemasa, I. and Tomomatsu, Y. and Hayashi, K. and Takagi, T.},
  booktitle={2018 International Conference on Information and Communications Technology},
  pages={226--233},
  year={2018},
  doi={10.1109/ICOIACT.2018.8350761}
}

@inproceedings{zhou2020interactive,
  title={Interactive recommender system via knowledge graph-enhanced reinforcement learning},
  author={Zhou, S. and Dai, X. and Chen, H. and Zhang, W. and Ren, K. and Tang, R. and Yu, Y.},
  booktitle={Proceedings of the 43rd ACM SIGIR Conference},
  pages={179--188},
  year={2020},
  doi={10.1145/3397271.3401174}
}

@article{fu2022deep,
  title={A deep reinforcement learning recommender system with multiple policies for recommendations},
  author={Fu, M. and Huang, L. and Rao, A. and Irissappane, A. A. and Zhang, J. and Qu, H.},
  journal={IEEE Transactions on Industrial Informatics},
  volume={19},
  number={2},
  pages={2049--2061},
  year={2022},
  doi={10.1109/TII.2022.3209290}
}

@article{zavadskas2016development,
  title={Development of TOPSIS method to solve complicated decision-making problems},
  author={Zavadskas, E. K. and Mardani, A. and Turskis, Z. and Jusoh, A. and Nor, K. M.},
  journal={International Journal of Information Technology and Decision Making},
  volume={15},
  number={3},
  pages={645--682},
  year={2016},
  doi={10.1142/S0219622016300019}
}

@article{shani2005mdp,
  title={An MDP-based recommender system},
  author={Shani, G. and Heckerman, D. and Brafman, R. I.},
  journal={Journal of Machine Learning Research},
  volume={6},
  pages={1265--1295},
  year={2005}
}

@inproceedings{panda2018topsis,
  title={TOPSIS in multi-criteria decision making: a survey},
  author={Panda, M. and Jagadev, A. K.},
  booktitle={2018 International Conference on Data Science and Business Analytics},
  pages={51--54},
  year={2018},
  doi={10.1109/ICDSBA.2018.00017}
}

@inproceedings{bangari2021review,
  title={A review on reinforcement learning based news recommendation systems and its challenges},
  author={Bangari, S. and Nayak, S. and Patel, L. and Rashmi, K. T.},
  booktitle={2021 International Conference on Artificial Intelligence and Smart Systems},
  pages={260--265},
  year={2021},
  doi={10.1109/ICAIS50930.2021.9395812}
}

@article{zheng2022survey,
  title={A survey of recommender systems with multi-objective optimization},
  author={Zheng, Y. and Wang, D. X.},
  journal={Neurocomputing},
  volume={474},
  pages={141--153},
  year={2022},
  doi={10.1016/j.neucom.2021.11.041}
}

@article{mukhametzyanov2021specific,
  title        = {Specific Character of Objective Methods for Determining Weights of Criteria in {MCDM} Problems: Entropy, {CRITIC} and {SD}},
  author       = {Mukhametzyanov, Irik Z.},
  journal      = {Decision Making: Applications in Management and Engineering},
  volume       = {4},
  number       = {2},
  pages        = {76--105},
  year         = {2021},
  issn         = {2560-6018},
  doi          = {10.31181/dmame210402076i}
}

@inproceedings{bied2023toward,
  author    = {Bied, Guillaume and Nathan, Solal and P{\'e}renn{\`e}s, Elia and Deleu, Johannes and Duflo, Guillaume and Piwowarski, Benjamin and Soulier, Laure},
  title     = {Toward Job Recommendation for All},
  booktitle = {Proceedings of the Thirty-Second International Joint Conference on Artificial Intelligence},
  year      = {2023},
  pages     = {5906--5914},
  doi       = {10.24963/ijcai.2023/655}
}

@article{gugnani2020skills,
  author  = {Gugnani, Akshay and Misra, Hemant},
  title   = {Implicit Skills Extraction Using Document Embedding and Its Use in Job Recommendation},
  journal = {Proceedings of the AAAI Conference on Artificial Intelligence},
  year    = {2020},
  volume  = {34},
  number  = {8},
  pages   = {13286--13293},
  doi     = {10.1609/aaai.v34i08.7038}
}

@article{reusens2017explicit,
  author  = {Reusens, Michael and Lemahieu, Wilfried and Baesens, Bart},
  title   = {A Note on Explicit Versus Implicit Information for Job Recommendation},
  journal = {Decision Support Systems},
  year    = {2017},
  volume  = {98},
  pages   = {26--35},
  doi     = {10.1016/j.dss.2017.04.002}
}

@article{dhameliya2019job,
  author  = {Dhameliya, Jignesh and Desai, Nirav P.},
  title   = {Job Recommendation System Using Content and Collaborative Filtering Based Techniques},
  journal = {International Journal of Soft Computing and Engineering},
  year    = {2019},
  volume  = {9},
  number  = {3},
  pages   = {8--13},
  doi     = {10.35940/ijsce.c3266.099319}
}

@article{senger2024karrierewege,
  author  = {Senger, Elena and Campbell, Yuri and van der Goot, Rob and Plank, Barbara},
  title   = {Karrierewege: A Large Scale Career Path Prediction Dataset},
  journal = {arXiv preprint arXiv:2412.14612},
  year    = {2024},
  doi     = {10.48550/arXiv.2412.14612}
}

@article{senger2025realistic,
  author  = {Senger, Elena and Campbell, Yuri and van der Goot, Rob and Plank, Barbara},
  title   = {Toward More Realistic Career Path Prediction: Evaluation and Methods},
  journal = {Frontiers in Big Data},
  year    = {2025},
  volume  = {8},
  pages   = {1564521},
  doi     = {10.3389/fdata.2025.1564521}
}

@article{alsaif2022matchedjob,
  author  = {Alsaif, Suleiman Ali and Hidri, Minyar Sassi and Eleraky, Hassan Ahmed},
  title   = {Learning-Based Matched Representation System for Job Recommendation},
  journal = {Computers},
  year    = {2022},
  volume  = {11},
  number  = {11},
  pages   = {161},
  doi     = {10.3390/computers11110161}
}

@article{channabasamma2022talent,
  author  = {Channabasamma, A. and Suresh, Yeresime},
  title   = {A Recommendation-Based Contextual Model for Talent Acquisition},
  journal = {Journal of Computer Science},
  year    = {2022},
  volume  = {18},
  number  = {7},
  pages   = {612--621},
  doi     = {10.3844/jcssp.2022.612.621}
}

@article{poncio2024systematicjob,
  author  = {Poncio, Flordeliza P.},
  title   = {Navigating Techniques in Job Recommender Systems on Internship Profile Matching: A Systematic Review},
  journal = {Journal of Research in Innovative Teaching \& Learning},
  year    = {2024},
  volume  = {17},
  number  = {2},
  pages   = {352--367},
  doi     = {10.1108/JRIT-01-2024-0016}
}

@article{zou2024reviewjob,
  author  = {Zou, Zhou and Huspi, Sharin Hazlin and Nuar, Ahmad Najmi Amerhaider},
  title   = {A Review on Job Recommendation System},
  journal = {Journal of Advanced Research in Applied Sciences and Engineering Technology},
  year    = {2024},
  volume  = {41},
  number  = {2},
  pages   = {113--124},
  doi     = {10.37934/araset.41.2.113124}
}

@article{du2024llmjob,
  author  = {Du, Yingpeng and Luo, Di and Yan, Rui and Li, Zhenkun and Ma, Yuanrun and Liu, Yang and Zhang, Ruobing and Zhang, Xu and Zhang, Leyu},
  title   = {Enhancing Job Recommendation through LLM-Based Generative Adversarial Networks},
  journal = {Proceedings of the AAAI Conference on Artificial Intelligence},
  year    = {2024},
  volume  = {38},
  number  = {8},
  pages   = {8363--8371},
  doi     = {10.1609/aaai.v38i8.28678}
}

@article{wu2024llmgraphjob,
  author  = {Wu, Likang and Qiu, Zhaopeng and Zheng, Zhi and Li, Yongqi and Xu, Han and Chen, Enhong and Xiong, Hui},
  title   = {Exploring Large Language Model for Graph Data Understanding in Online Job Recommendations},
  journal = {Proceedings of the AAAI Conference on Artificial Intelligence},
  year    = {2024},
  volume  = {38},
  number  = {8},
  pages   = {9178--9186},
  doi     = {10.1609/aaai.v38i8.28769}
}

@inproceedings{schellingerhout2024explainablejob,
  author    = {Schellingerhout, Roan},
  title     = {Explainable Multi-Stakeholder Job Recommender Systems},
  booktitle = {Proceedings of the 18th ACM Conference on Recommender Systems},
  year      = {2024},
  pages     = {1318--1322},
  doi       = {10.1145/3640457.3688014}
}

@article{sun2021skillrl,
  author  = {Sun, Ying and Zhuang, Fuzhen and Zhu, Hengshu and Zhang, Qi and Xiong, Hui and He, Qing},
  title   = {Cost-Effective and Interpretable Job Skill Recommendation with Deep Reinforcement Learning},
  journal = {Proceedings of The Web Conference 2021},
  year    = {2021},
  pages   = {3827--3838},
  doi     = {10.1145/3442381.3449985}
}

@article{sun2025marketaware,
  author  = {Sun, Ying and Ji, Yang and Zhu, Hengshu and Zhou, Xiangyu and Zhang, Qi and Xiong, Hui},
  title   = {Market-Aware Long-Term Job Skill Recommendation with Explainable Deep Reinforcement Learning},
  journal = {ACM Transactions on Information Systems},
  year    = {2025},
  volume  = {43},
  number  = {2},
  pages   = {1--35},
  doi     = {10.1145/3704998}
}

@article{azri2025contextjob,
  author  = {Azri, Muhammad and Haw, Su-Cheng and Ng, Kok-Why},
  title   = {Context-Aware Job Recommender System},
  journal = {JOIV: International Journal on Informatics Visualization},
  year    = {2025},
  volume  = {9},
  number  = {2},
  pages   = {877},
  doi     = {10.62527/joiv.9.2.3021}
}

@article{tang2025explainablepjr,
  author  = {Tang, Fang and Zhu, Renqi and Yao, Feng and Li, Qiang and Guo, Xiaofei},
  title   = {Explainable Person--Job Recommendations: Challenges, Approaches, and Comparative Analysis},
  journal = {Frontiers in Artificial Intelligence},
  year    = {2025},
  volume  = {8},
  pages   = {1660548},
  doi     = {10.3389/frai.2025.1660548}
}

@article{zhang2020explainable,
author  = {Zhang, Yongfeng and Chen, Xu},
title   = {Explainable Recommendation: A Survey and New Perspectives},
journal = {Foundations and Trends in Information Retrieval},
year    = {2020},
volume  = {14},
number  = {1},
pages   = {1--101},
doi     = {10.1561/1500000066}
}

@article{hidasi2015gru4rec,
  author  = {Hidasi, Bal{\'a}zs and Karatzoglou, Alexandros and Baltrunas, Linas and others},
  title   = {Session-based Recommendations with Recurrent Neural Networks},
  journal = {arXiv},
  year    = {2015},
  pages   = {arXiv:1511.06939},
  doi     = {10.48550/arXiv.1511.06939}
}

@inproceedings{kang2018sasrec,
  author    = {Kang, Wang-Cheng and McAuley, Julian},
  title     = {Self-Attentive Sequential Recommendation},
  booktitle = {2018 IEEE International Conference on Data Mining (ICDM)},
  year      = {2018},
  pages     = {197--206},
  doi       = {10.1109/ICDM.2018.00035}
}

@article{cureton1956rankbiserial,
  author  = {Cureton, Edward E.},
  title   = {Rank-Biserial Correlation},
  journal = {Psychometrika},
  year    = {1956},
  volume  = {21},
  number  = {3},
  pages   = {287--290},
  doi     = {10.1007/BF02289138}
}

@inproceedings{johary2025jobhop,
  author    = {Johary, Iman and Romero, Rapha\"el and Mara, Alexandru and De Bie, Tijl},
  title     = {{JobHop}: A Large-Scale Dataset of Career Trajectories},
  booktitle = {Proceedings of the 2025 IEEE International Conference on Big Data (BigData)},
  year      = {2025},
  publisher = {IEEE},
	pages   = {2184--2191},
  doi       = {10.1109/BigData66926.2025.11402454}
}

@article{guan2024jobformer,
  author  = {Guan, Zhihao and Yang, Jia-Qi and Yang, Yang and Zhu, Hengshu and Li, Wenjie and Xiong, Hui},
  title   = {{JobFormer}: Skill-Aware Job Recommendation with Semantic-Enhanced Transformer},
  journal = {ACM Transactions on Knowledge Discovery from Data},
  year    = {2024},
  doi     = {10.1145/3701735}
}

@article{zha2024careermobility,
  author  = {Zha, Rui and Qin, Chuan and Zhang, Le and Shen, Dazhong and Xu, Tong and Zhu, Hengshu and Chen, Enhong},
  title   = {Career Mobility Analysis With Uncertainty-Aware Graph Autoencoders: A Job Title Transition Perspective},
  journal = {IEEE Transactions on Computational Social Systems},
  year    = {2024},
  doi     = {10.1109/TCSS.2023.3239038}
}

@article{cui2026scp,
  author  = {Cui, Shuai and Sun, Yunhe and Zhang, Yan and Meng, Qing and Zhu, Hengshu},
  title   = {{LLM}-Enhanced Career Knowledge Graph Understanding for Job Mobility Prediction},
  journal = {ACM Transactions on Management Information Systems},
  year    = {2026},
  doi     = {10.1145/3787466}
}

@article{nematollahi2026maudrl,
  author  = {Nematollahi, M. and Guitouni, A. and Izadyar, N. and Belacel, N. and Park, A.},
  title   = {Multi-Attribute Utility Deep Reinforcement Learning Method for Sequential Multi-Criteria Decision Problems: Application to Human Resource Planning},
  journal = {Computers and Operations Research},
  year    = {2026},
  doi     = {10.1016/j.cor.2026.107426}
}

@article{hoque2026fuzzytopsis,
  author  = {Hoque, Shahria and Karim, Asif and Alam, Mahmudul and Gope, Niladri},
  title   = {When {LLM} Meets {Fuzzy}-{TOPSIS} for Personnel Selection Through Automated Profile Analysis},
  journal = {IEEE Access},
  year    = {2026},
  volume  = {14},
  pages   = {15778--15794},
  doi     = {10.1109/ACCESS.2026.3658575}
}

@article{feng2026personjob,
  author  = {Feng, J. and Yang, J. and Li, S. and Miao, Q. and Xi, Y. and Xia, Z.},
  title   = {Enhancing Person-Job Fit Through Multi-Temporal Career Trajectory Modeling},
  journal = {Expert Systems with Applications},
  year    = {2026},
  volume  = {300},
  pages   = {130413},
  doi     = {10.1016/j.eswa.2025.130413}
}

@article{liu2026cfrlhybrid,
  author  = {Liu, Haowei and Ismail, F. and Zhang, W. and Zou, P. and Hussain, T. and Sharma, Y. and Lilhore, U. and Simaiya, S. and Tekeste, L.},
  title   = {A Scalable Hybrid Framework for Boosting Customer Experience and Operational Efficiency in E-Commerce},
  journal = {Scientific Reports},
  year    = {2026},
  doi     = {10.1038/s41598-026-37437-7}
}

\end{document}